\documentclass[%
 reprint,
superscriptaddress,
 amsmath,amssymb,
 aps,
 prf,
 onecolumn,
 longbibliography,
]{revtex4-2}

\usepackage{graphicx}
\usepackage{dcolumn}
\usepackage{bm}

\usepackage{gensymb}
\usepackage[utf8]{inputenc}
\usepackage{hyperref}
\usepackage{color}
\usepackage{ulem}
\usepackage{soul}
\usepackage{todonotes}
\newcommand{\dif}{\mathrm{d}}

\usepackage{siunitx}

\begin{document}

\preprint{APS/123-QED}

\title{Contact-angle hysteresis provides resistance to drainage of liquid-infused surfaces in turbulent flows}

\author{Sofia Saoncella}
\affiliation{FLOW, Dept. of Engineering Mechanics, Royal Institute of Technology  (KTH),  100 44 Stockholm, Sweden}%
\author{Si Suo}
\affiliation{FLOW, Dept. of Engineering Mechanics, Royal Institute of Technology  (KTH),  100 44 Stockholm, Sweden}%
\author{Johan Sundin}
\affiliation{FLOW, Dept. of Engineering Mechanics, Royal Institute of Technology  (KTH),  100 44 Stockholm, Sweden}%
\author{Agastya Parikh}
\affiliation{Institute of Fluid Mechanics and Aerodynamics, Universität der Bundeswehr München, Neubiberg, 85577, Germany}%
\author{Marcus Hultmark}
\affiliation{Mechanical and Aerospace Engineering Department, Princeton University, Princeton, NJ 08544, USA}%
\author{Wouter Metsola van der Wijngaart}
\affiliation{Division of micro and nanosystems, KTH, Royal Institute of Technology, 10044 Stockholm, Sweden.}
\author{Fredrik Lundell}
\affiliation{FLOW, Dept. of Engineering Mechanics, Royal Institute of Technology  (KTH),  100 44 Stockholm, Sweden}%
\author{Shervin Bagheri}
\affiliation{FLOW, Dept. of Engineering Mechanics, Royal Institute of Technology  (KTH),  100 44 Stockholm, Sweden}%

\date{\today}

\begin{abstract}
Lubricated textured surfaces immersed in liquid flows offer tremendous potential for reducing fluid drag, enhancing heat and mass transfer, and preventing fouling. According to current design rules, the lubricant must chemically match the surface to remain robustly trapped within the texture. However, achieving such chemical compatibility poses a significant challenge for large-scale flow systems, as it demands advanced surface treatments or severely limits the range of viable lubricants. In addition, chemically tuned surfaces often degrade over time in harsh environments. Here, we demonstrate that a lubricant-infused surface (LIS) can resist drainage in the presence of external shear flow without requiring chemical compatibility. Surfaces featuring longitudinal grooves can retain up to 50\% of partially wetting lubricants in fully developed turbulent flows. The retention relies on contact-angle hysteresis, where triple-phase contact lines are pinned to substrate heterogeneities, creating capillary resistance that prevents lubricant depletion.  We develop an analytical model to predict the maximum length of pinned lubricant droplets in microgrooves. This model, validated through a combination of experiments and numerical simulations, can be used to design chemistry-free LISs for applications where the external environment is continuously flowing. Our findings open up new possibilities for using functional surfaces to control transport processes in large systems.
\end{abstract}

\maketitle


\section{Introduction}
Functional surfaces that have the ability to regulate mass, energy and momentum transport in a fluctuating fluid flow could significantly reduce energy waste in various applications, including marine infrastructure, medical devices, thermal systems and food processing units. One promising technology that has emerged is lubricant-infused surfaces (LISs), which use microstructures to trap a lubricating liquid \cite{Wong2011, Lafuma2011} \cite{Wong2011, Lafuma2011}. The presence of a lubricant offers several ways to interact favourably with external fluid flows. The interface between the lubricant and the overlying liquid is slippery, preventing the attachment of microorganisms and resulting in an efficient anti-fouling surface \cite{Epstein2012,zouaghi2017,charpentier2015}. Additionally, the flow over a LIS experiences slippage, reducing the frictional resistance exerted on the surface \cite{Solomon2014,pakzad2022,fu2017,van2017,lee2019}. Moreover, the external flow can set the trapped lubricant into recirculating motion, enhancing heat and mass transfer rates between the surface and the bulk flow \cite{Sundin2022}.

Previous studies on submerged LISs have  relied on a chemical compatibility rule that was primarily developed for liquid-repellent and anti-adhesive applications \cite{Lafuma2011, Wong2011, Smith2013, Preston2017}. According to this rule, the chemistry of the surface needs to be similar to the chemistry of the lubricant to avoid dewetting.  However, achieving chemical compatibility poses a significant challenge for large-scale applications. Most techniques used to tune solid surface energy involve spray coating or thin film deposition \cite{Villegas2019, Li2019, Baumli2021}, which can be time-consuming and costly when applied to large surface areas. In addition, chemically modified surfaces  degrade  over time due to exposure to UV light, humidity, chemicals or stresses during  handling.  An alternative approach of achieving chemical compatibility is  selecting a lubricant and a solid with a high affinity. However, such combinations are limited and based on hydrophobic  polymers and other inert lubricants that raise environmental concerns as their inertness makes them difficult to degrade naturally \cite{Peppou-Chapman2020}.

In this study, we characterize the physics of LISs in the presence of an external fluid flow  when the chemical compatibility rule is broken. We find that the substrate can retain a significant amount of lubricant by relying on contact-angle hysteresis (CAH). Small-scale physical or chemical inhomogeneities naturally appearing on the substrate pin the lubricant-liquid-solid contact line and create a capillary force that resists lubricant depletion. The size of the pinning force does not depend on the equilibrium contact angle $\theta_e$, but instead on the difference between the advancing ($\theta_{\mathrm{adv}}$) and receding ($\theta_{\mathrm{rec}}$) contact angles. This implies that the condition that the lubricant must preferentially wet the surface rather than the overlying liquid is not strictly necessary for a LIS submerged in liquid that flows. We develop a theoretical model that provides an explicit expression for \textit{a priori} predicting the maximum length of a lubricant droplet given the surface geometry, CAH and external friction force. Our criterion applies in turbulent flow environments, showing that the retention mechanism is robust under unsteady and fluctuating external stresses. Moreover, we use the model to introduce a non-equilibrium design rule for chemistry-free LISs. The rule provides novel opportunities for manufacturing lubricant-infused surfaces to control transport processes in large-scale flow systems.

\section{Experimental configuration and methods}
\subsection{Turbulent channel flow facility }\label{subsec:tubulent flow}
\begin{figure}[t]
\centering
\includegraphics[width=0.9\linewidth]{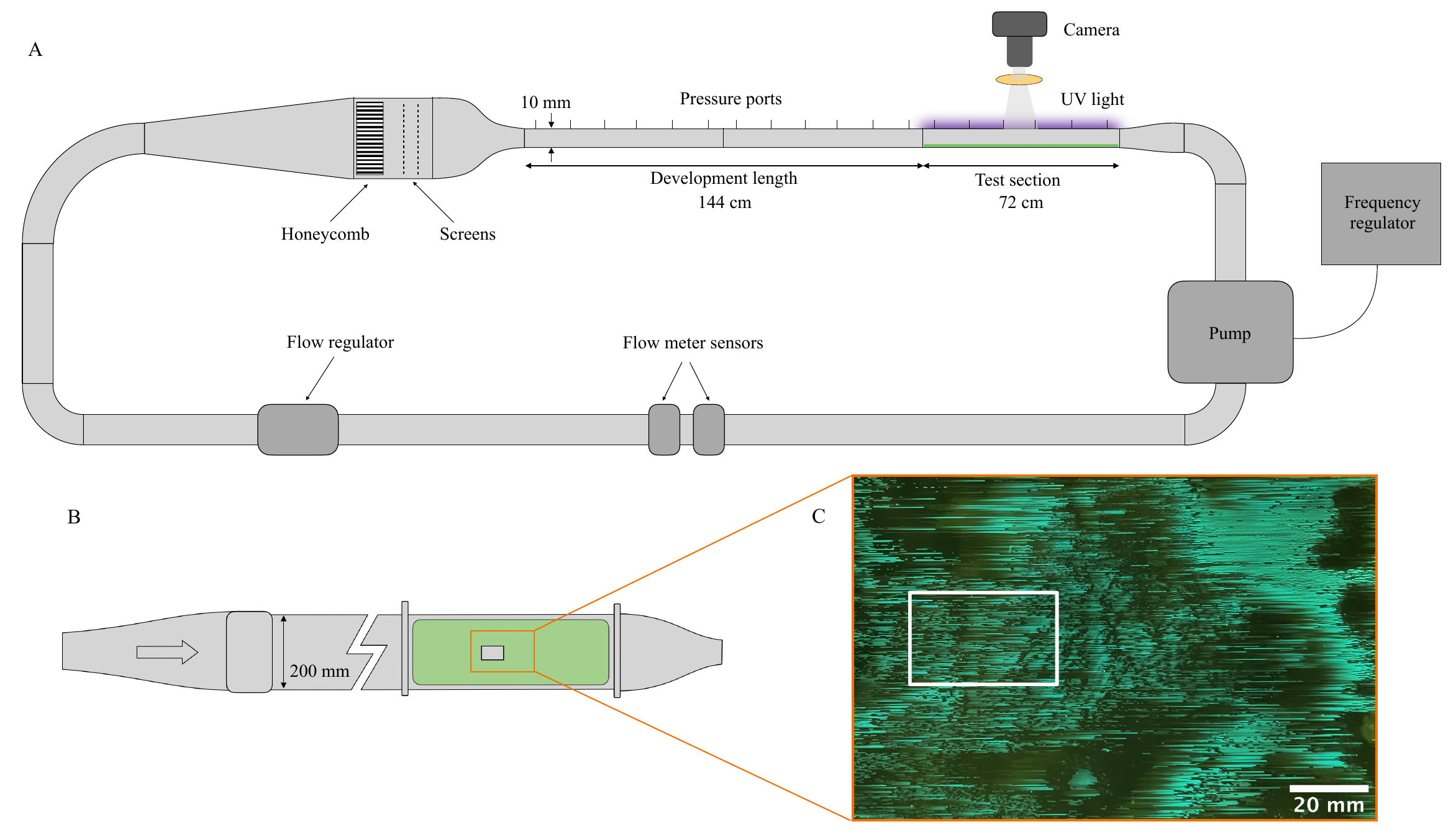}
\caption{
(A) Schematic representation of F-SHARC (Fluid-Surface-High-Aspect-Ratio-Channel) flow facility. The flow is generated by a \SI{11}{\kilo\watt} centrifugal pump which is operated by a frequency controller.
The flow rate is measured by means of an ultrasonic flowmeter (OMEGA FDT-25 W) with an accuracy of $\pm 1\%$. Each section of the channel has six pressure taps with a diameter of \SI{0.4}{\milli\meter} and spaced by \SI{10}{\centi\meter}, positioned along the centreline. The pressure gradient $dp/dx$ along the length of the test section is measured with a series of piezoelectric miniature transducers (Honeywell ABPDRRV001PDAA5) with an accuracy of $\pm0.25\%$ of full scale.
(B) Schematic top view of the channel test section showing the area covered by the LIS and the camera field of view (FOV) (2.5$\times$3.7 cm$^2$) represented by the grey rectangle.
(C) Image of a portion of LIS (13$\times$9 cm$^2$) after being exposed  to a maximum wall-shear stress (WSS) $\tau_{\max}=\SI{10.4}{\pascal}$ for two hours. The white rectangle corresponds to the camera FOV.}
\label{fig:Channel}
\end{figure}

We use a water channel facility to characterize the behavior of liquid-infused surfaces in turbulence. The flow facility, shown schematically in Fig.~\ref{fig:Channel}A, has the width  $w_{ch}=\SI{200}{\milli\meter}$ and the height $h_{ch}=\SI{10}{\milli\meter}$, resulting in an aspect-ratio of 20:1. 
The flow is developed over a length of $144h_{ch}$ before reaching the test section which measures $72h_{ch}$ in length. 
%
%
%
%
For a fully developed turbulent channel flow, which has a mean velocity that is uniform in the spanwise and streamwise directions, the mean wall shear stress (WSS) $\tau_w$, can be obtained from the force balance,
%
%
\begin{equation}
    \tau_w=-\frac{h_{ch}}{2}\frac{dp}{dx}.
    \label{eq:WSS}
\end{equation}
Here, $dp/dx$ represents the mean pressure gradient, which is obtained by measuring the difference between each pressure port and the most upstream one with a series of piezoelectric miniature transducers (see Fig.~\ref{fig:Channel}).  Tab. \ref{tab:flow_cond} presents the mean WSS obtained from \eqref{eq:WSS} on a smooth solid wall for three different bulk velocities ($U_b$) . The bulk velocities range from $0.91$ m/s to $1.56$ m/s, resulting in mean WSS from $3.2$ Pa to $7.8$ Pa. These values correspond to bulk Reynolds number $Re=U_bh_{ch}/\nu$ (with $\nu$ being the fluid viscosity) ranging from 9100 to 15600 and skin friction coefficient $C_f=\tau_w/(0.5\rho U_b^2)$ (where $\rho$ is the fluid density) from $7.73\cdot10^{-3}$ to $6.41\cdot10^{-3}$. 
To investigate if the facility generates a canonical turbulent channel flow, we compare the $C_f$ obtained from measurements with the empirical relation proposed by Dean \cite{dean1978reynolds}, given by $C^*_f=0.073Re^{-0.25}$. Tab. \ref{tab:flow_cond}  demonstrates a sufficiently good agreement between the skin-friction coefficients for the purposes of our investigation. In a similar way, Tab. \ref{tab:flow_cond} compares the measured friction Reynolds number, $Re_\tau=u_\tau h_{ch}/(2\nu)$ -- ranging from 280 to 442 --  with the empirical relation $Re^*_{\tau} = 0.166Re^{0.88}$, where again a reasonable agreement is observed. The main sources of error are attributed to the accuracy of the pressure sensors and to temperature variations, which were seen to vary by approximately one degree over five minutes of flow at the highest flow rate.
%
\begin{table}[tbhp]
  \begin{center}
  \caption{Specifications for the three turbulent flow configurations. The bulk velocity ($U_b$) and the mean turbulent wall shear stress ($\tau_w=-h_{ch}\Delta P/(2L)$) were measured using an ultrasonic flowmeter and piezoelectric pressure transducers, respectively. The bulk Reynolds number ($Re$) and the friction Reynolds number ($Re_\tau$) can be calculated based on the measurements and the channel height ($h_{ch}=10$ mm), kinematic viscosity of water ($\nu=10^{-6}$ m$^2$/s) and the friction velocity ($u_\tau=\sqrt{\tau_w/\rho}$). The friction Reynolds number  can also be calculated using empirical relation ($Re^*_\tau$). The agreement between $Re_\tau$ and $Re^*_\tau$ serves as a validation that the experimental facility leverages a canonical turbulent channel flow.
  }
  \vspace{5mm}
  \begin{tabular}{llccc}
  \hline\vspace{1mm}
  \textbf{Quantity} &  \textbf{Acquisition/Expression} &  \textbf{Case 1} &  \textbf{Case 2} &  \textbf{Case 3}  \\\hline\vspace{1mm}
       \vspace{1mm}
       $U_b$ (m/s) & Ultrasonic flowmeter & 0.91 $\pm$ 0.01 & 1.30 $\pm$ 0.01 & 1.56 $\pm$ 0.02\\\vspace{1mm}
       $\tau_w$ (Pa) & Pressure transducers & 3.2 $\pm$ 0.7 & 5.8 $\pm$ 0.8 & 7.8 $\pm$ 1.0 \\\vspace{1mm}
       $Re$ & $U_b h_{ch}/\nu$ & 9100 & 13000 & 15600\\\vspace{1mm}
       $C_f\times 10^{3}$ & $\tau_w/(1/2\rho U_b^2)$ & 7.73 & 6.86 & 6.41 \\\vspace{1mm}
       $C^*_f\times 10^{3}$ & $\approx 0.073Re^{-0.25}$ & 7.47 & 6.84 & 6.53 \\\vspace{1mm}
       $Re_{\tau}$ & $u_\tau h_{ch}/(2\nu)$ & 283 & 381 & 442\\\vspace{1mm}
       $Re^*_{\tau}$ & $ 0.166Re^{0.88}$ & 275 & 376 & 442\\
       \hline
  \end{tabular}
  \label{tab:flow_cond}
  \end{center}
\end{table}

\subsection{Liquid-infused surfaces}\label{chem_compat}
\begin{figure}[t]
\centering
\includegraphics[width=0.8\linewidth]{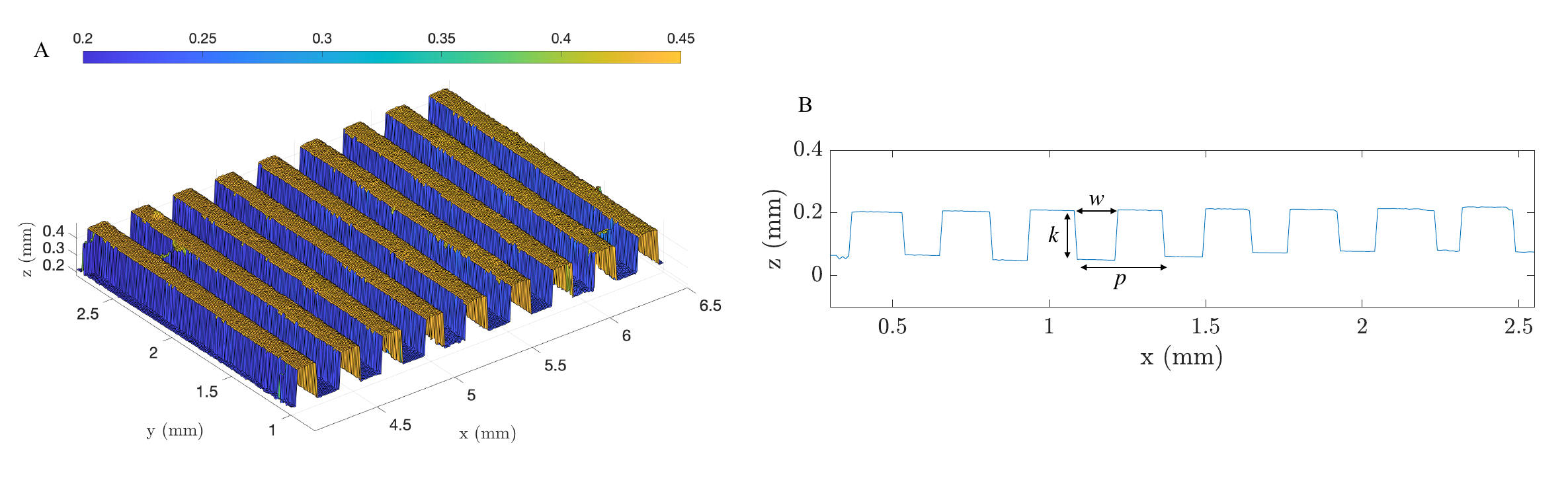}
\caption{
(A) Height map of the textured substrate used for the LISs extracted from a tomographic scan. (B) Cross section profile of the same sample. 
%
The three-dimensional scan is obtained using an optical coherence tomography (OCT) device (Thorlabs, Telesto II) with an axial resolution in water of $\SI{2.58}{\micro\meter}.$
}
\label{fig:OCT-scan}
\end{figure}

%
The liquid-infused surface is flush-mounted to the lower wall of the test section. The texture consists of longitudinal (i.e.~parallel to the flow direction) grooves
fabricated using UV-lithography (described in App.~\ref{app:fabrication}).
Fig.~\ref{fig:OCT-scan} shows a portion of the groove geometry and the corresponding profile.
The grooves have depth $k=149\pm22 \mu$m, width $w=143\pm5 \mu$m and pitch $p=276\pm2\mu$m.
\noindent Before mounting the panel with the substrate on the channel wall, the grooves are infused with hexadecane (Sigma-Aldrich) (viscosity ratio with respect to water $\mu_l/\mu_w=3.7$ and lubricant-water surface tension $\gamma=\SI{53}{\milli\newton/\meter}$). The surfaces are then slightly tilted for 5 minutes to drain the excess lubricant by gravity. The final sample, composed by four adjacent substrate tiles, covers the full area of the test section wall with an extension of $15\times60~\unit{\cm}^2$. 
The upper wall is made of a smooth clear acrylic plate that allows for optical access. To visualise the lubricant-water interface during the flow experiments, a fluorescent imaging technique is used. The lubricant fluid is mixed with a fluorescent dye (Tracer Products TP-4300) at a volume ratio 2:1000. A series of UV LED lights, fixed to the test section, excite the dye which emits a green glow. Fig.~\ref{fig:Channel}C shows how the lubricant appears during a flow experiment. Using a digital camera (Nikon D7100 DSLR) with a Nikon AF Micro-Nikkor $\SI{200}{\milli\meter}$ lens and a yellow filter, consecutive images are taken at a selected time interval. The resolution of the photos is 160 px/mm. The field of view (FOV) of the camera is $375\times 25$ mm$^2$ and it is located at the centre of the test section, as represented in Fig.~\ref{fig:Channel}B. The duration of the measurements is two hours for each experiment.

The surface chemistry together with the substrate geometry determines the spontaneous spreading of the lubricant in the textured surface in the presence of water. For our solid substrate, which has the geometry of a rectangular groove with width $w$ and depth $k$, the surface energy per unit length due to a displacement $dx$ is given by \cite{Quere2008}, 
\begin{equation*}
    \text{d}E = \gamma_{so}(2k+w)\text{d}x+\gamma_{ol}w\text{d}x-\gamma_{sl}(2k+w)\text{d}x,
\end{equation*}
where $\gamma_{so}, \gamma_{ol}, \gamma_{sl}$ are the surface tensions between the different phases: lubricant droplet ($o$), solid substrate ($s$), and the immiscible surrounding liquid ($l$). The lubricant will wick into the groove when it is energetically favourable ($dE<0$), i.e.
\begin{equation*}
    (\gamma_{so}-\gamma_{sl})(2k+w)+\gamma_{ol}w< 0.
\end{equation*}
By defining spreading parameter as $S=\gamma_{sl}-\gamma_{so} - \gamma_{ol}$, we can write the condition above,
\begin{equation}
    S>-\gamma_{ol}\frac{2k}{2k+w}.
    \label{eq:SI-spreeading}
\end{equation}
When the spreading condition (\ref{eq:SI-spreeading}) is satisfied, a lubricant droplet will spread in the groove until a thin film is formed. In this scenario, for which the lubricant and the solid are chemically compatible, the LIS is called \textit{wetting}.
On the other hand, when \eqref{eq:SI-spreeading} is not satisfied, the droplet will only partially wet the grooved surface, and a triple-phase contact line appears. In the latter configuration, the lubricant-solid combination is chemically incompatible and the LIS is \textit{partially wetting}.

Tab.~\ref{tab:contact angles} shows the equilibrium chemical properties of the partially wetting LIS and the wetting LIS that we have used in this study. The former has untreated polymeric substrate (see App. \ref{app:fabrication}), whereas the latter surface is functionalised with a hydrophobic coating.
We measured the equilibrium contact angle ($\theta_e$) of a lubricant droplet immersed in water on smooth samples with the sessile drop method \cite{eral2013contact}.  We observe that for partially wetting LIS, the smooth surface prefers water over hexadecane ($\theta_e>90^\circ$), while for the wetting LIS, we have the opposite situation ($\theta_e<90^\circ$). 
The advancing and receding contact angles ($\theta_{adv}$ and $\theta_{rec}$), were measured by extruding and withdrawing liquid into and from the lubricant droplet (for details of the measurements see App~\ref{app:fabrication}).  The contact-angle hysteresis, defined as $\Delta\theta=\theta_{adv}-\theta_{rec}$, is much larger for the partially wetting surface compared to the wetting one. Cured polymeric surfaces, such as PDMS, are known to have a very large $\Delta\theta$.

\vspace{5mm}
\begin{table}[tbhp]
\centering
\caption{Measured values of equilibrium ($\theta_e$), advancing ($\theta_\textrm{adv}$), receding  ($\theta_{rec}$) contact angles and contact-angle hysteresis ($\Delta\theta$) of the partially wetting and wetting substrates.}
\begin{tabular}{lccccc}
\hline
LIS &  Eq.\eqref{eq:SI-spreeading} & $\theta_e$ & $\theta_{adv}$ & $\theta_{rec}$ & $\Delta\theta$ \\
\hline
Partially wetting & Not satisfied & $117^\circ\pm9^\circ$ & $134^\circ\pm9^\circ$ & $12^\circ\pm2^\circ$ & $122^\circ\pm10^\circ$ \\
Wetting &  Satisfied &$42^\circ\pm4^\circ$ & $62^\circ\pm8^\circ$ & $16^\circ\pm5^\circ$ & $46^\circ\pm9^\circ$ \\
\hline
\end{tabular}
\label{tab:contact angles}
\end{table}

%


\section{Retention of partially wetting lubricant in turbulence}
%
%
\begin{figure*}[tbhp]
\centering
\includegraphics[width=1\linewidth]{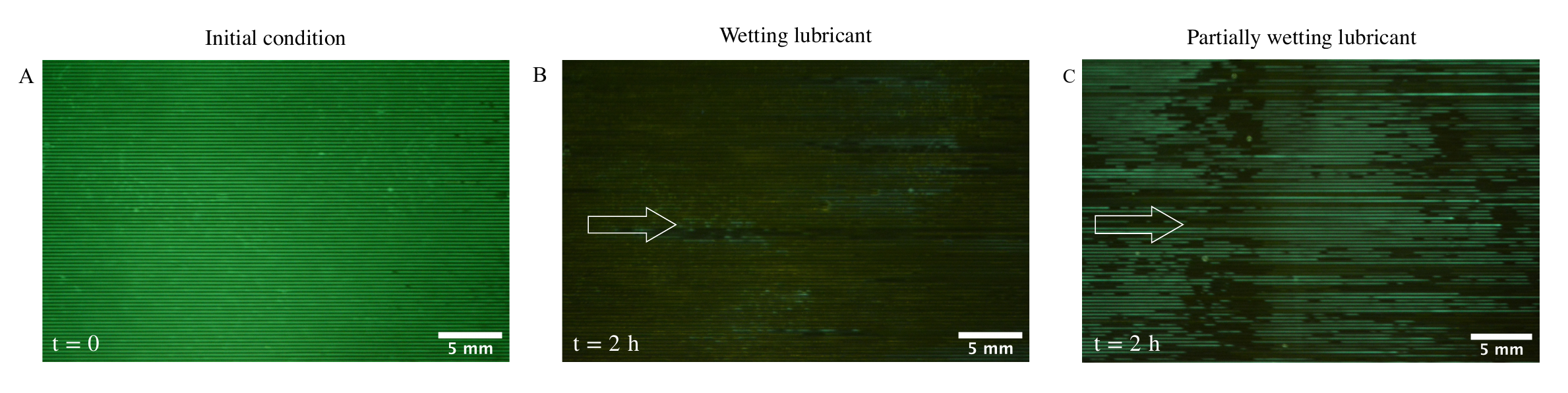}
\caption{Observations of wetting and partially wetting LIS in a turbulent flow. 
%
(A) The LIS at the beginning of the experiment where, in wetting and partially wetting configuration, the grooves are completely filled with lubricant (in green). 
(B) The grooves of a wetting LIS at maximum WSS $\tau_{\max}=\SI{5.8}{\pascal}$ are  drained from the lubricant after two hours. 
(C) The grooves of a partially wetting LIS retain $46\%$ of the lubricant after two hours.
}
\label{fig:schematic_experiments}
\end{figure*}
%

We start by characterizing the lubricant-water interface for the flow configuration Re=9100 (Tab. \ref{tab:flow_cond}) in the presence of the partially wetting and wetting LISs (Tab. \ref{tab:contact angles}).
Figure \ref{fig:schematic_experiments}A shows the wetting LIS at the beginning of the experiment where the grooves are filled with lubricant (green). The initial state of the partially wetting LIS looks exactly the same. We observed different states of two LISs after approximately 2 hours exposure to the turbulent flow. The wetting LIS -- which adheres to the established design principles where the lubricant fully wets the substrate -- was drained, leaving behind only a thin film at the bottom of the grooves where the flow velocity is very small (Fig.~\ref{fig:schematic_experiments}B).  Lubricant depletion in longitudinal grooves without barriers in a shear flow is expected \cite{Wexler2015}. 
For partially-wetting LIS -- as illustrated in Fig.~\ref{fig:schematic_experiments}C -- after two hours, 46\% of the initial lubricant volume remained in the grooves. This demonstrates that a partially wetting lubricant can resist drainage when exposed to a flow. Somewhat contradictory, the chemically incompatible LIS, which breaks the equilibrium design rule offers a greater resistance to drainage in non-equilibrium conditions. Appendix \ref{app:lubricant_drainage} provides a more quantitative assesment of the lubricant drainage for different configurations.

\begin{figure*}[tbhp]
\centering
\includegraphics[width=0.8\linewidth]{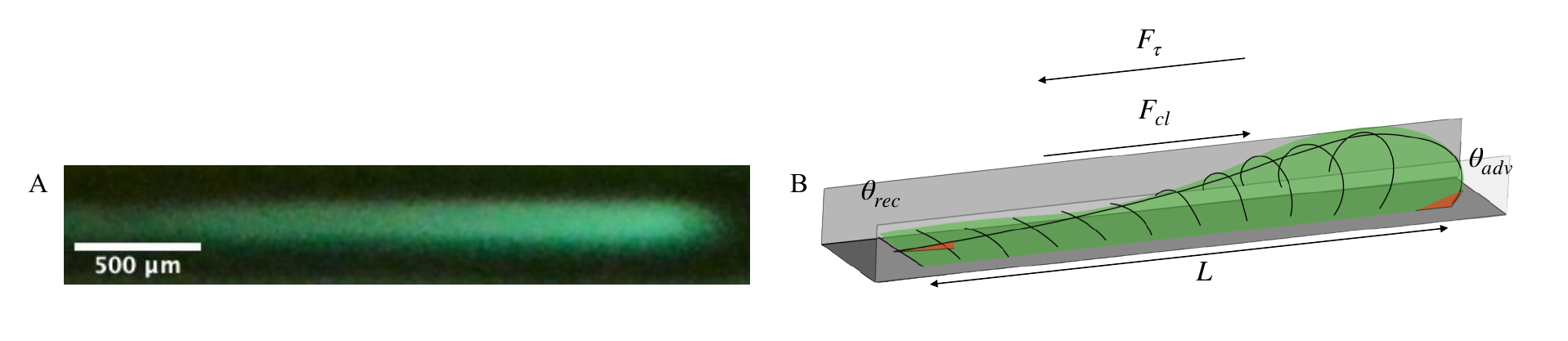}
\caption{
(A) A lubricant droplet pinned in the groove of a partially wetting LIS.
(B) Sketch of the interface shape of lubricant within the groove showing the advancing $\theta_{\mathrm{adv}}$ and receding $\theta_{\mathrm{rec}}$ contact angles, the lubricant length $L$, the external shear stress $F_{\tau}$ and the pinning force $F_{cl}$.
}
\label{fig:lub_drop}
\end{figure*}

The experiments (SI Movies S2, S3, and S4) of partially wetting LISs reveal that the lubricant-water interface initially breaks up, allowing water to infiltrate the grooves, after which the lubricant forms elongated droplets pinned to the surface (Fig.~\ref{fig:lub_drop}A). At the contact lines, where the water-lubricant-solid phases meet, there is an adhesive force that depends on the physical and chemical heterogeneity of the substrate. The force can be expressed as  \cite{Furmidge1962}, 
\[
F_{cl} \sim  w \gamma (\cos \theta_{\mathrm{rec}} - \cos \theta_{\mathrm{adv}}),
\]
as represented in Fig.~\ref{fig:lub_drop}B. This adhesive force acts in the opposite direction to the external hydrodynamic force, 
\[
F_{\tau}\sim \tau_s L w,
\]
where $\tau_s$ is the (averaged) shear stress on the lubricant-liquid interface, and $L$ is the length of the lubricant droplet. 
By equating the two forces, we can define the largest possible length  of a stationary droplet as 
\[L_\infty \sim {\gamma}/{\tau_s} (\cos \theta_{\mathrm{rec}} - \cos \theta_{\mathrm{adv}}).\]
This means that, in the presence of triple-phase contact lines, the adhesive force can resist the imposed shear stress for lubricant droplets of length $L<L_\infty$. Droplets longer than $L_\infty$, on the other hand, will be displaced because the shear stress exceeds the pinning force.
%
%
As we will quantitatively demonstrate in the following sections, the partially wetting LIS has pinned contact lines and sufficiently large CAH to yield stationary lubricant droplets reaching up to 35 mm in length. In contrast, the wetting LIS spreads into a thin film at the bottom of the grooves without well-defined contact lines.

\section{Theoretical  model of maximum retention length}
\begin{figure}[tbhp]
\centering
\includegraphics[width=0.8\linewidth]{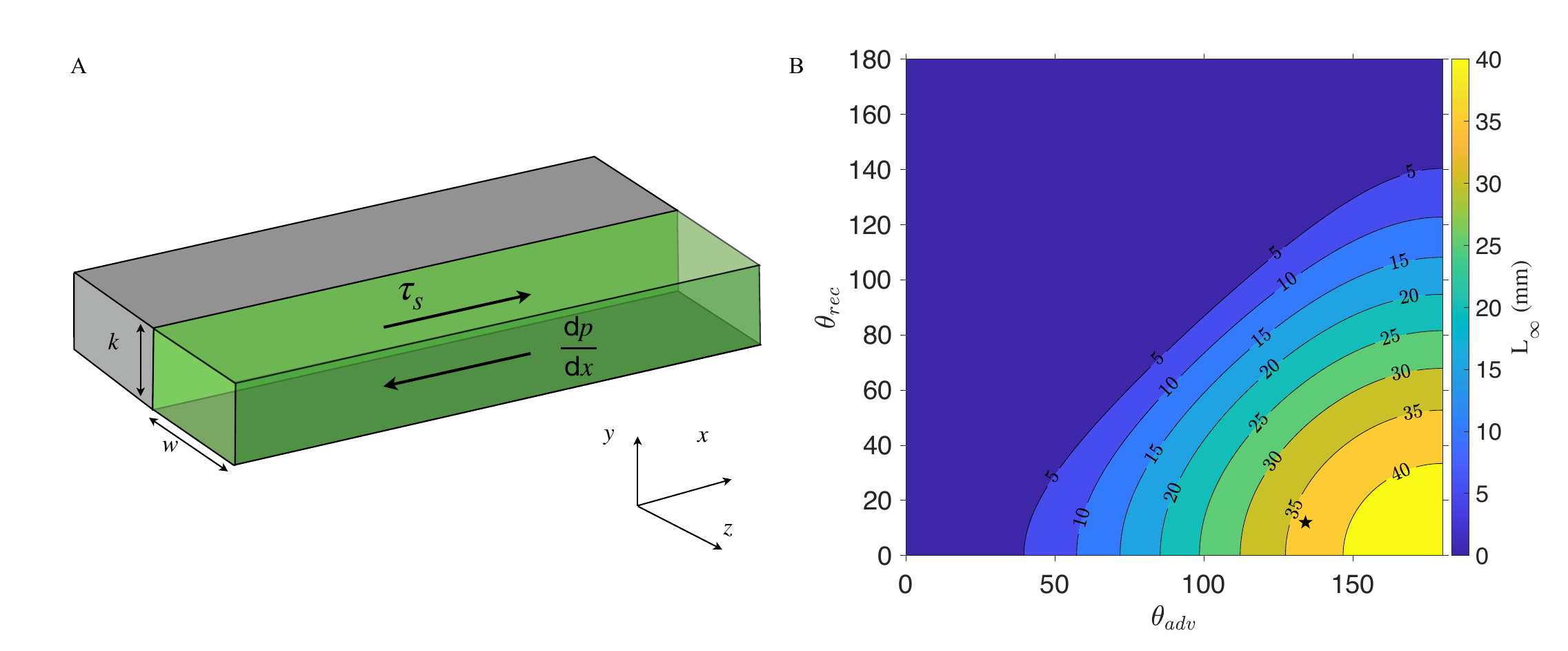}
\caption{(A) The idealised configuration of a groove segment filled with the lubricant used to develop the analytical model. (B) The maximum retention length  predicted by the model at shear stress $\tau_s=\SI{5.8}{\pascal}$ for different advancing and receding contact angles combinations. The star indicates the present experimental conditions, where $\theta_{\mathrm{adv}}=134^\circ \pm 9^\circ$ and $\theta_{\mathrm{rec}}=12^\circ \pm 2^\circ$, which results in $L_\infty=35.8$ mm.}
\label{fig:analytics_setup_varplot}
\end{figure}
Having observed significant lubricant retention of partially wetting LIS in turbulent flows, we now develop a model that predicts the maximum droplet length for a given shear stress and contact-angle hysteresis.  We consider a long groove that is filled with a lubricant and subjected to a uniform  interface shear stress $\tau_{s}$ (Fig.~\ref{fig:analytics_setup_varplot}A). 
The imposed viscous stress generates a steady and unidirectional lubricant flow in the streamwise direction resulting in a flux given by \cite{Wexler2015},
\[
q_{\tau} = c_s\frac{wk^2\tau_s}{\mu_{l}},
\]
where $c_\mathrm{s}$ is a geometry-dependent constant (see App.~\ref{app:analytics}) and $\mu_l$ is the lubricant viscosity. For stable retention, there must be a flux in the opposite direction that exactly balances $q_{\tau}$. Assuming that the opposite lubricant flow is driven by a constant pressure gradient $\dif p/\dif x$, the associated flux can be written as \cite{Wexler2015},
\[
q_{p} = -c_p\frac{wk^3}{\mu_l}\frac{\dif p}{\dif x},
\]
where $c_\mathrm{p}$ is another geometry-dependent constant. The balance of the two fluxes results in,
\begin{equation}
    \frac{\dif p}{\dif x} = \frac{\tau_s}{k}\frac{c_s}{c_p}.
    \label{eq:balancingPressure:2}
\end{equation}
The pressure gradient arises from the adhesion force at the contact line of a droplet in the groove.
The force \cite{deGennes2004} is defined as
\begin{equation}
    F_{cl} = \oint_\textit{cl} \gamma \left (\mathbf{n}_{cl}\cdot \mathbf{e}_x \right ) \dif l, 
\end{equation}
where $\mathbf{n}_{cl}$ is a unit vector that is tangent to the liquid-lubricant interface and normal to the contact line. The integration is performed along a closed curve, where the lubricant-liquid-solid phases meet (App.~\ref{app:analytics}). We assume a constant angle along the entire contact line of upstream and downstream parts of the droplet, given by $\theta = \theta_{\mathrm{adv}}$ and $\theta = \theta_{\mathrm{rec}}$, respectively. Under these assumptions and since the droplet completely wets the side walls of the groove, we obtain 
\begin{equation}
    F_{cl} = -(w + 2k)\gamma(\cos \theta_\textrm{rec} - \cos \theta_\textrm{adv}).
\end{equation}
With $F_{\mathrm{cl}}$ assumed to change the pressure equally over the projected area $wk$ in the $yz$-plane, we can write,
\begin{equation}
    \frac{\dif p}{\dif x} = -\frac{F_{cl}}{Lwk},
    \label{eq:PressureGradientAdhesiveForce}
\end{equation}
where $L$ is the length of the droplet. By inserting \eqref{eq:PressureGradientAdhesiveForce} into the flux balance (\ref{eq:balancingPressure:2}), we can define the maximum retention length as,
\begin{equation}
    L_{\infty} = A\frac{\gamma}{\tau_s}\left (\cos \theta_{\mathrm{rec}} - \cos \theta_{\mathrm{adv}}\right ).
    \label{eq:maximumRetentionLength:2}
\end{equation}
The constant $A$ depends on the substrate geometry, which for longitudinal grooves is given by  
\[
A=\frac{w + 2k}{w}\frac{c_p}{c_s}.
\] 
Droplets of length $L$ are stationary if 
\begin{equation}
L \leq L_\infty,
    \label{eq:stableLength}
\end{equation}
as they can induce a sufficient pressure gradient to balance the imposed external shear stress. Droplets with $L > L_\infty$, on the other hand, will move downstream in the groove.
%
%

Fig.~\ref{fig:analytics_setup_varplot}B shows how $L_\infty$ varies with the advancing and receding contact angles for $\tau_s=5.8$ Pa. We observe that  retention lengths of a few millimetres can be obtained from relatively small CAH ($\Delta \theta \approx 30^\circ$). Almost all surfaces exhibit some degree of contact angle hysteresis \cite{Samuel2011, Laroche2023}.  Highly smooth surfaces, such as those made of PTFE (polytetrafluoroethylene), typically exhibit low contact angle hysteresis ($\Delta \theta=10^\circ$), while PDMS (polydimethylsiloxane) and Rose-petal type of surfaces can have $\Delta \theta$ exceeding $100^\circ$ \cite{allred2019, Bhushan2010}.
The model developed herein contains the essential  ingredients for the CAH-based retention mechanism. We have neglected inertial and gravitational effects in the lubricant and assumed a large viscosity ratio ($\mu_l/\mu_w \gg 1$, which indicates negligible slippage at the lubricant-liquid interface \cite{Liu2016}). Furthermore, long-range forces such as van der Waals forces have been neglected, which implies the absence of  nanometric precursor films in the grooves. 

Previous work \citep{Wexler2015, Wexler2015B,fu2019} have, instead of CAH, introduced physical or chemical barriers in the grooves to generate a shear-resisting pressure gradient. The maximum retention length 
provided by Wexler \textit{et.~al.}~\cite{Wexler2015} for a lubricant-infused longitudinal groove that terminates in a lubricant reservoir reads,
%
\begin{equation}
    L_{\infty,W} = \frac{k}{r_{\min}}\frac{\gamma}{\tau_s},
    \label{eq:maximumRetentionLength-wexlerT}
\end{equation}
where $r_{\min}$  approximates the interface curvature at the tail of the droplet. Equation \eqref{eq:maximumRetentionLength-wexlerT} is a special case of  \eqref{eq:maximumRetentionLength:2} with $\theta_{\mathrm{adv}}=90^\circ$, since the interface curvature is a consequence of the pinning force.

\section{Retention distribution and numerical simulations}

\begin{table}[t]
\centering
\caption{
Three experiments of partially wetting LIS were performed at different bulk flow velocities ($U_b$), where $\tau_{\max}$ is the corresponding maximum streamwise wall-shear stress. Also reported are the maximum retention lengths predicted by the analytical model $L_{\infty}$, computed with $c_s=0.07$, $c_p=0.05$ (obtained by solving Stokes equations for the given microgroove geometry), $\theta_{\mathrm{adv}}=134^\circ$, $\theta_{\mathrm{rec}}=12^\circ$ and $\tau_s=\tau_{\max}$. The theoretical values can be compared to the maximum retention lengths measured experimentally  $L_{\mathrm{max}}$. Last column reports the percentage of drained lubricant after two hours (see App.~\ref{app:lubricant_drainage} for definition of $\% LD$).
}
\begin{tabular}{cccccccc}
\hline
$Re$ & $U_b$ $(\si{m/s})$ & $\tau_{\max}$ $(\si{\pascal})$ & $L_{\infty}$ $(\si{\milli\meter})$ & $L_{\mathrm{max}}$ $(\si{\milli\meter})$ & $\% \mathrm{LD}$ \\
\hline
9100 & 0.9 & 5.8 &  35.8 & 32.2 $\pm$ 0.2 & 54\\
13000 & 1.3 & 10.4 & 19.8 & 14.6 $\pm$ 0.2 & 71\\
15600 & 1.6 & 14.0 & 14.7 & 12.2 $\pm$ 0.2 & 82\\
\hline
\end{tabular}
\label{tab:retention_lenghts}
\end{table}
To verify that stationary droplets satisfy \eqref{eq:stableLength}, we now consider  water channel experiments at higher flow velocities (Re=13000 and Re=15600) using the partially wetting LIS (Tab.~\ref{tab:retention_lenghts}).
%
In a turbulent flow, the streamwise wall-shear stress (WSS) near the lubricant-liquid interface fluctuates, resulting in intermittent peaks that can be up to 80\% higher than the mean wall-shear stress \cite{alfredsson1988, cheng2020}. 
Since the time scale of the fluctuations is on the order of milliseconds, we expect that all lubricant-liquid interfaces will be exposed to the peak WSS within a 2-hour interval.  Consequently, the pinning force must be sufficiently large to withstand the maximum WSS.
Therefore, when evaluating \eqref{eq:maximumRetentionLength:2} for turbulent flows, the shear stress $\tau_s$ is approximated by  $\tau_{\max}=\max(\tau_w+\tau^\prime_x)$,
where  $\tau^\prime_x $ is the streamwise WSS fluctuations. 
The maximum WSS can be estimated from, 
\[\tau_{\max} \approx \tau_w+2\tau^\prime_{x,rms}
\approx\tau_w+2(0.4 \tau_w)
\approx 1.8\tau_w, \]
where $\tau^\prime_{x,rms}$ is the root-mean-square of the streamwise WSS. The approximations above are based on relations and PDFs obtained from experiments  \cite{alfredsson1988} and numerical simulations \cite{cheng2020}.
%
%
%
Using $\tau_s=\tau_{\max}$ in \eqref{eq:maximumRetentionLength:2}, we calculated the predicted maximum retention lengths $L_{\infty}$ which range from 35.8 mm for the lowest flow speed to 14.7 mm for the fastest flow speed (Tab.~\ref{tab:retention_lenghts}).
The corresponding measured maximum retention lengths, $L_{\max}$, range from 32.2 mm to 12.2 mm. The agreement with the predictions is remarkable, considering that we have not used empirical fitting parameters despite dealing with a turbulent flow.

\begin{figure}
\centering
\includegraphics[width=0.9\linewidth]{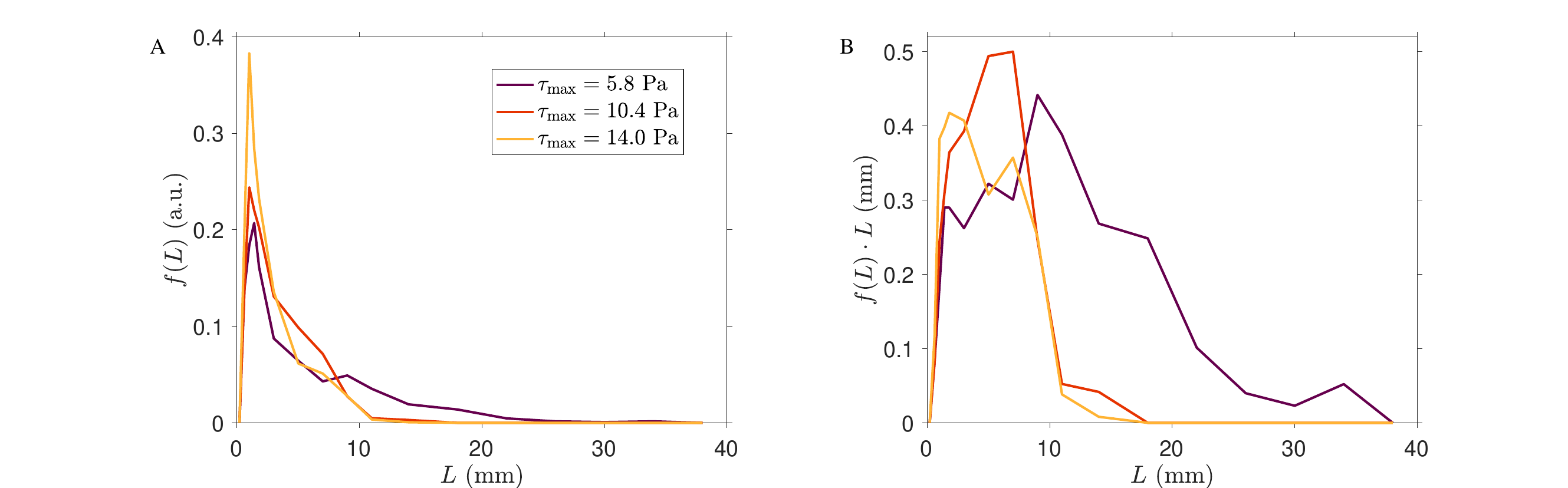}
\caption{(A) Probability density function $f(L)$ of droplet lengths for different shear stresses. (B) Distribution multiplied by the droplet length, which shows the contribution of droplets of different lengths to the total lubricant volume.}
\label{fig:lenghts_distributions}
\end{figure}

Lubricant droplets with maximum lengths are however rare. The stationary LIS has many smaller droplets that can generate higher pressure gradients for opposing the shear stress (as indicated by Eq.~\ref{eq:PressureGradientAdhesiveForce}). To quantify the distribution of droplet lengths in the stationary state, we calculated the probability density function $f(L)$ (Fig.~\ref{fig:lenghts_distributions}A). We observe that droplets with a length of approximately 1.5 mm appear most frequently across all shear stress values. Droplets with lengths greater than 10 mm have a low occurrence rate, particularly at high shear stresses.
While short droplets dominate in quantity, longer droplets contribute significantly to overall lubricant retention. The total volume of lubricant retained in the stationary LIS can be approximated by $V\approx wkN\langle L\rangle$, where $wk$ is the constant cross-sectional area of the groove, $N$ the number of retained droplets in the surface and $\langle L\rangle$ is the mean lubricant length, which is given by 
\begin{equation}
\langle L \rangle = \int_0^{\infty} f(L)\cdot L\  \textrm{d}L.
\label{eq:V}
\end{equation}
This indicates that the total lubricant volume depends on the product, $f(L)\cdot L$, shown in Fig.~\ref{fig:lenghts_distributions}(B).  The contribution to $V$ is fairly equal from droplets with lengths ranging from $1$ mm to $10$ mm for two cases of high shear stress, and with lengths between $1$ mm and $20$ mm for the lower shear stress case. This implies that short and long droplets contribute equally to the volume of lubricant that is retained by the surface. 

\begin{figure*}
\centering
\includegraphics[width=1\linewidth]{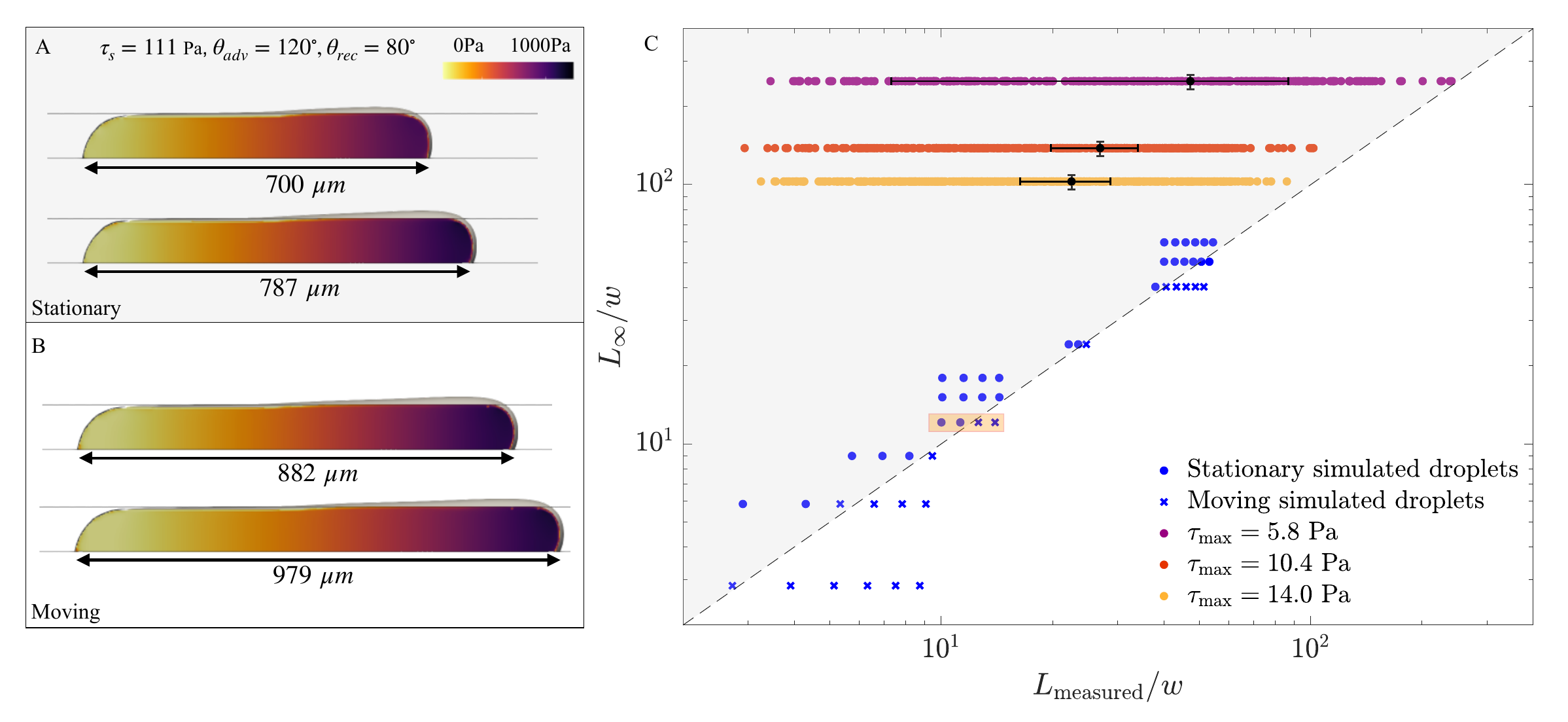}
\caption{Side view of the simulated lubricant droplets in one groove in the stationary regime ($L<L_\infty$) (A) and moving regime ($L>L_\infty$) (B). The fluid flows from left to right. The head of the drop is slightly larger than the groove, as shown by the grey shadow. The colour bar represents the pressure inside the drop. (C) Retention map showing the normalized retention lengths measured from experiments and numerical simulations on the horizontal axis and the normalized retention lengths predicted by theory on the vertical axis. The theoretical lengths are obtained by inserting the shear stress, CAH, groove width, height and surface tension that corresponds to each experiment and simulation into \eqref{eq:maximumRetentionLength}.
For the configurations that fall into the grey upper left region, the measured lengths are shorter than the model prediction, thus in a stable regime.  All the lengths measured in the experiments of partially wetting LIS after two hours  (purple, red and yellow symbols) fall in this region as they do not move. The black symbol and horizontal error bars indicate the mean values and variances of the lengths distributions. The vertical error bars indicate the uncertainty in the measurement of the advancing and receding contact angles. 
The simulated lubricant droplets (blue symbols), which are too long to withstand the imposed shear stress (blue crosses) fall in the bottom right region of the graph. Shorter simulated droplets remain pinned to the substrate and can balance the shear stress (blue circles).  The four symbols in the orange box correspond to the  simulated drops shown in frames A and B. The parameters of the other simulations are given in App.~\ref{app_numerics}}.
\label{fig:retention_map}
\end{figure*}

The theoretical retention length $L_\infty$ is the upper limit of a  range of droplet lengths that exist in turbulence. To demonstrate that the theoretical model is predictive when excluding the turbulent fluctuations, we have performed a set of numerical simulations using the open-source software OpenFOAM \cite{roenby2016computational}. See Appendix \ref{app_numerics} for details of the numerical method. We considered the laminar shear (Couette) flow over a single lubricant-infused groove, with the same density and viscosity for the infused and external fluids.  We modelled CAH by pinning the contact line when the apparent contact angle is within the hysteresis window $\left[\theta_{\mathrm{rec}}, \  \theta_{\mathrm{adv}} \right]$; otherwise, the contact line moves with a prescribed constant contact angle.  The computational domain had dimensions $\left(L_x, \ L_y, \ L_z\right)=\left(20k, \ w, \ 2k\right)$ with a groove size ratio $w/k \approx 1$, similar to our experiments. We imposed a constant $\tau_s$ and $\Delta \theta$, and then gradually increased the volume of a lubricant droplet in the groove.

Figs.~\ref{fig:retention_map}A and B show how the lubricant droplet is deformed by the shear stress $\tau_s=\SI{111}{\pascal}$ when $\Delta \theta=40^\circ$. The pressure difference $\Delta P$ between the front and back end of the droplet is induced by contact-line force that pins the interface to the surface, causing the interface away from the surface to deform. The downstream and upstream contact lines dominate the adhesion force, so $\Delta P$ is the same for droplets of all sizes. However, the pressure gradient ($\sim \Delta P /L $) is only sufficiently strong to withstand the imposed shear stress when the droplet length is smaller than a critical value.
Our numerical simulations show that the droplet is stationary below $L= 787$ $\mu$m, but is moving when $L=882$ $\mu$m (Figs.~\ref{fig:retention_map}A and B). The maximum possible droplet length obtained from \eqref{eq:maximumRetentionLength:2}  is $L_\infty=800$ $\mu$m, which lies in between the stationary and the moving droplets.


The numerical and experimental results are collected in a retention map in Fig.~\ref{fig:retention_map}C. 
The map shows the predicted retention lengths normalized by the groove width, $L_{\infty}/w$, plotted against the measured retention lengths, $L_{\mathrm{measured}}/w$.   
The dashed line represents $L_{\mathrm{measured}}=L_\infty$, the gray region marks $L_{\mathrm{measured}}<L_\infty$ (stationary droplets) while the white region marks $L_{\mathrm{measured}}>L_\infty$ (moving droplets).  The latter case indicates that the resistance due to contact line hysteresis is unable to withstand the applied shear stress.  All the measured lubricant droplets in the turbulent channel flow experiments  -- evaluated after 2 hours -- fall into the stable grey region of the map. The large filled markers represent the mean of the distributions of the retention lengths, and the horizontal error bars represent their variance. We note that the mean and maximum values of the distributions decrease with a slope similar to the stability boundary. This indicates consistency in the scaling of the retention lengths with the shear stress.
The numerical simulations allow varying the shear stress and CAH over a wide range ($\tau_s\in [33,111]$ Pa and $\Delta \theta \in [10^{\circ},60^{\circ}]$). We observe a quantitative agreement between the numerical simulations and the theory, where nearly all the stationary (circles) and moving (crosses) lubricant droplets fall into the stable and unstable regimes, respectively. This also confirms that the retention length in unidirectional flows is quantitatively given by \eqref{eq:maximumRetentionLength:2} while the expression serves as an upper limit in the presence of turbulence. 

\section{Discussion}
For liquid-repellent and anti-adhesive applications, where the external environment is near equilibrium, LISs rely on chemical compatibility to keep the lubricant robustly trapped in the microstructures. Expression \eqref{eq:SI-spreeading} determines lubricant-solid  compatibility specifically for lubricant spreading in  streamwise grooves  \cite{Quere2008}, but can be generalized to any surface texture \cite{Preston2017}.
%
Condition (\ref{eq:SI-spreeading}) is often fulfilled by tuning the surface chemistry, for example, by depositing a layer of octadecyltrichlorosilane (OTS) onto silicon surfaces \cite{Smith2013}.  

\begin{figure}[tbhp]
\centering
\includegraphics[width=0.8\linewidth]{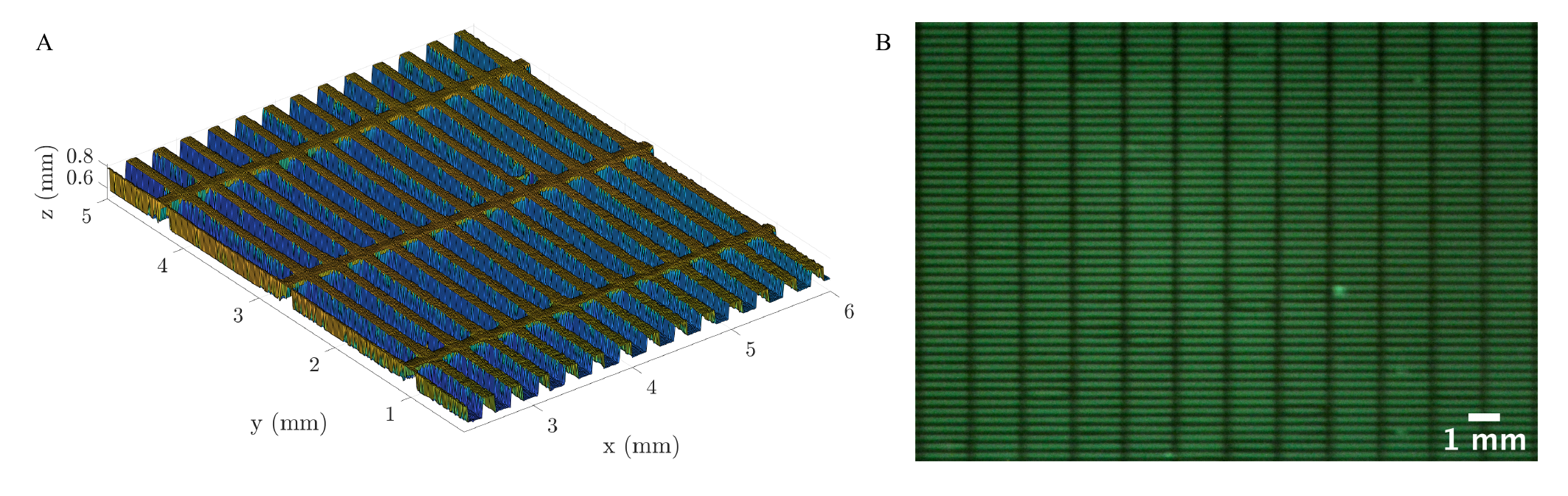}
\caption{Partially wetting LIS with physical barriers. (A) Three-dimensional scan of the solid substrate consisting of periodic cavities. The cross-sectional dimension of the cavities is the same as in the longitudinal grooves, whereas the cavity length is 150 mm. The thickness of the barriers is 140 $\mu$m. (B) Lubricant remains in the cavities after being subjected to a turbulent flow with maximum WSS $\tau_{\max}=\SI{5.8}{\pascal}$ for two hours.}
\label{fig:cavity_lis}
\end{figure}

Our findings provide a starting point for developing chemistry-free lubricant-infused surfaces for applications where the external environment is continuously flowing. In Fig.~\ref{fig:cavity_lis}A, we present a LIS design featuring cavities of length $L_{\mathrm{cavity}}=$1.5 mm and with the same width and height as the previous longitudinal grooves. The length of the cavities correspond to half of the average drop length measured at the flow case 3. For the partially wetting lubricant, we obtained $>80\%$ retention (Fig.~\ref{fig:cavity_lis}B) after 2 hours in the presence of turbulent flow ($\tau_{\max}=\SI{5.8}{\pascal}$). The $20\%$ depletion was due to the initial dewetting (Fig.~\ref{fig:perc_lubricant_drainage}).
This demonstrates that a LIS, where \eqref{eq:SI-spreeading}  is not satisfied, can stably trap lubricant droplets if 
\begin{equation}
L_{\mathrm{cavity}} < L_\infty.
\label{eq:stableLengthCavity}
\end{equation}
To provide a design criterion for a general microstructured surface, we define a capillary number as $Ca=L_{\mathrm{cavity}}\tau_s/\gamma$, which represents the balance between the external viscous force and the  capillary force within a surface cavity. The condition in (\ref{eq:stableLengthCavity}) can be recast as a non-dimensional criterion for a drop to remain pinned in the microstructured surface,
\begin{equation}
 Ca<\Delta_\theta,
 \label{eq:critical_Capillary}
 \end{equation}
where
\begin{equation}
\Delta_\theta =A (\cos \theta_{\mathrm{rec}} - \cos \theta_{\mathrm{adv}}),
 \label{eq:wettingTransition}
\end{equation}
and
\begin{equation}
A = \frac{c_p}{c_s} \frac{1}{\langle w \rangle}\oint_{cl} \mathbf{ n}_{cl} \cdot \mathbf{e}_x\ dl.
 \label{eq:wettingTransitionA}
\end{equation}
The critical number, $\Delta_\theta$, is a function of the surface microstructure properties and independent of the flow. 
The coefficient $A$ in \eqref{eq:wettingTransitionA}  takes into account three essential surface features. The ratio $c_p/c_s$ -- which  can be obtained by solving Stokes equations in a unit cell of the texture -- accounts for the relative ease for which lubricant flux is generated in the texture  from  imposed pressure gradient ($c_p$) and shear stress ($c_s$).  Moreover, $A$ is inversely proportional to the streamwise-averaged groove width, ${\langle w \rangle}$, since narrower structures result in less exposed interfacial area to external shear stress. 
Finally, the integral in \eqref{eq:wettingTransitionA} represents the total length for which the contact line force has a component in the $x$-direction. An analogue condition to \eqref{eq:critical_Capillary} exists for the configuration where a drop partially wets an inclined surface. The condition reads $Bo<\Delta_\theta$, where $Bo=L\rho g/\gamma$ is the Bond number and gravity is the driving external force \cite{Furmidge1962}. 

As shown by Wexler \textit{et.~al.}~\cite{Wexler2015}, when both conditions  (\ref{eq:SI-spreeading}) and (\ref{eq:critical_Capillary}) are fulfilled, the lubricant spontaneously wets the textured surface and retention is enforced through distinct physical or chemical barriers that generate a resisting Laplace pressure gradient. Such self-healing LIS \cite{Wong2011, Lafuma2011} are useful in applications that require high precision control of multiple transport processes, such as microfluidic devices, batteries, microprocessors and micro-heat exchangers. 
We have shown  that significant retention in the presence of a flow can be achieved through a resistive pinning force without satisfying the equilibrium criterion (\ref{eq:SI-spreeading}). This comes, however,  at the cost of a loss (10\%-20\% in our experiments) of lubricant caused by a rapid initial dewetting process. In many large turbulent flow systems, partially lubricated surfaces may still offer significant functionality. Examples include marine systems, food processing units, medical devices and thermal systems. Chemistry-free LIS  would significantly increase the possible choices of lubricants for these applications by circumventing the difficulty in fabricating and sustaining large chemically-tuned surfaces. Additionally, chemistry-free LIS would address the environmental concern when using large volumes of hydrophobic polymers and other inert lubricants, which degrade very slowly in nature.

In summary, we have investigated a new retention mechanism of lubricant-infused surfaces that relies on contact-angle hysteresis of a substrate. We found that a partially wetting lubricant that naturally develops triple-phase contact lines can withstand relatively large shear stress in dynamic environments. We have derived an expression of the maximum possible  length of lubricant droplets, $L_\infty$, and validated the expression in laminar and turbulent flows. 
Our study offers the prospect of a new class of LIS for submerged conditions. This can contribute to developing large-scale lubricated surfaces that remain clean and energy-efficient in harsh flow environments.

\section*{Acknowledgements} We acknowledge support from Kunt and Alice Wallenberg foundation (KAW 2016.0255) and the Swedish Foundation for Strategic Research (FFL15:0001). We thank Susumu Yada and Saumey Jain for helping with surface fabrication. Access to the computational resources used for this work were provided by the Swedish National Infrastructure for Computing (SNIC).

\section*{Author contributions} Sofia Saoncella (SoSa) performed experiments, Johan Sundin (JS) developed the theory and So Suo (SiSu) performed numerical simulations. Agastya Partikh (AP), SoSa, Marcus Hultmark (MH) and Shervin Bagheri (SB) constructed the water channel facility. SoSa, SiSu, JS, Wouter Metsola van der Wijngaart (WmW) and  Fredrik Lundell (FL) and SB analyzed data. SoSa and SB wrote the paper with feedback from all authors.
\appendix

\section{Substrate fabrication and characterization}\label{app:fabrication}
The solid substrates of the LISs are fabricated with a soft lithographic method from Ostemer 322 (Mercene Labs, Stockholm, Sweden), Off-Stoichiometry-Thiol-Ene resin (OSTE). A flat layer of resin with size $150\times\SI{140}{\milli\meter}^2$ is cured on a black plastic sheet by exposure to UV radiation for $\SI{60}{\second}$. A second layer of resin is then prepared and cured for $\SI{60}{\second}$ with a geometric pattern by filtering the UV light through a photomask decorated with the desired 2D geometry, in this case, longitudinal grooves. The depth of the grooves is determined by the thickness of spacers on which the photomask rests.  The uncured resin is washed from the sample in an ultrasonic bath of PGMEA (Propylene-Glycol-Methyl-Ether-Acetate, Sigma-Aldrich) and dried with compressed air for three times. Finally, the surface is cured in an oven at $100^\circ$C for one hour. The hardened resin can then be coated to modify its surface energy. The low-energy substrate used for the partially wetting LIS was left uncoated, while the high-energy substrate, used for the wetting LIS, was spray-coated with a super water-repellent coating (HYDROBEAD-T). The final substrate is composed of four parts fabricated as described and mounted adjacently. 

\begin{figure}[tbhp]
\centering
\includegraphics[width=0.8\linewidth]{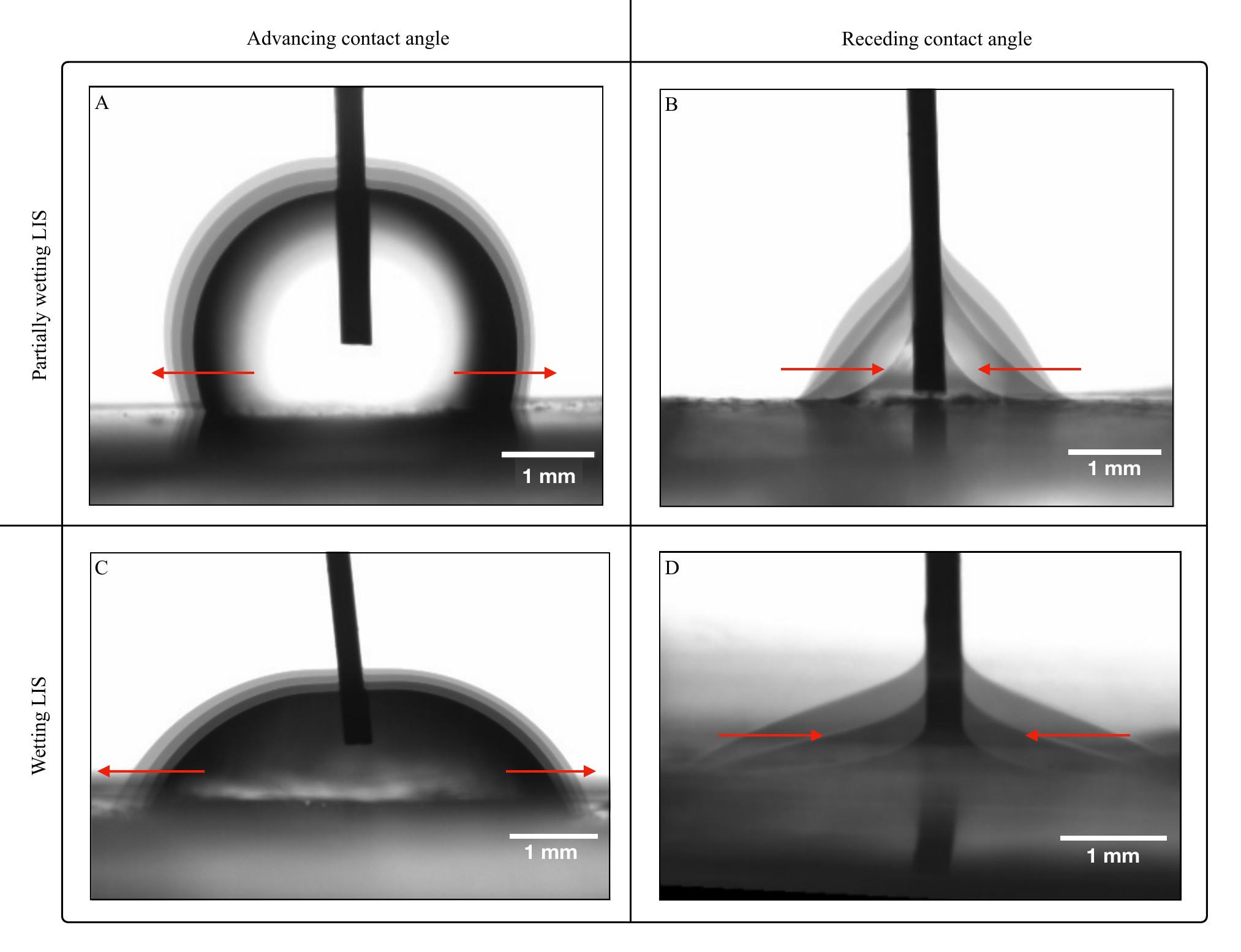}
\caption{Measurement of advancing and receding contact angles. The figures show a drop of hexadecane deposited on a smooth sample of the partially wetting substrate (A, B) and wetting substrate (C, D) while immersed in water. The images superimpose four instances in time during the lubricant infusion (A, C) and withdrawal (B, D) to show the evolution of the drop shape. The red arrows indicate the direction of the moving interface during the measurements.}
\label{fig:dynamic_ca}
\end{figure}

The contact angle of a drop of lubricant on the samples immersed in water are measured using the sessile drop method in an inverted setup configuration. Since the density of hexadecane is lower than that of water, the lubricant drop is generated from a thin needle (diameter $\SI{310}{\micro\meter}$, Hamilton, Gauge 30, point style 3) placed underneath the solid substrate immersed in water. As the drop is generated, buoyancy lifts it toward the sample surface, and the needle holding the drop is gradually brought closer to the surface until contact is made. The drop has a volume of about $\SI{5}{\micro\liter}$ and the oil is pumped or withdrawn by a syringe pump (New Era Pump Systems Inc., NE-4000) at a flow rate of $\SI{0.1}{\micro\liter/\second}$ to measure advancing and receding angles, respectively (Fig.~\ref{fig:dynamic_ca}). The angle right before the contact line starts to advance (recede) is defined as the advancing $\theta_{\mathrm{adv}}$ (receding $\theta_{\mathrm{rec}}$) contact angle. The contact angles measurements were repeated with ten drops for each case in different positions on the substrate; the reported angles correspond to the average values.

\section{Evaluation of lubricant drainage}\label{app:lubricant_drainage}
\begin{figure}[tbhp]
\centering
\includegraphics[width=0.9\linewidth]{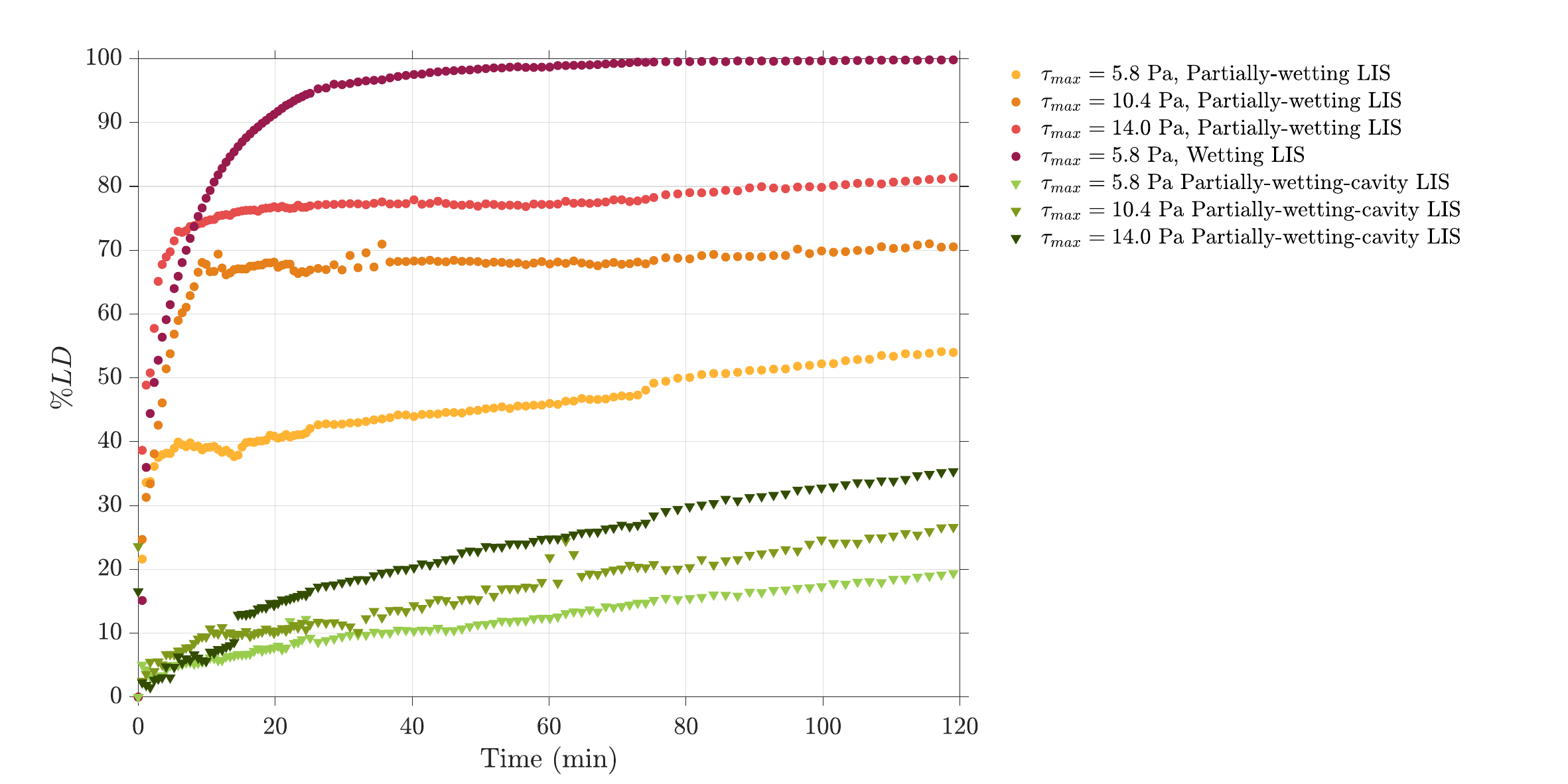}
\caption{
The graph depicts the percentage of lubricant drainage ($\%LD$) over time for all the investigated cases.}

\label{fig:perc_lubricant_drainage}
\end{figure}

The percentage of lubricant drainage with respect to the initial condition is defined as
\begin{equation*}
    \%\text{LD} = 100\cdot\left(1-\frac{h(t)}{h(t_0)} \right),
\end{equation*}
where $h(t)$ represents the volume of lubricant (per unit area) that infuses the grooves, evaluated at time $t$. In the expression above, $t_0$  denotes the initial time. 
The quantity $h(t)$ is extracted from the acquired images, assuming that the height of the lubricant in the groove is directly proportional to the pixel intensity. Therefore, the lubricant volume per unit area is computed as the sum of pixel intensities $I_i$ at time $t$, 
\begin{equation}
    h(t)=k\sum_{i} I_i(t).
\end{equation}
%
The temporal evolution of lubricant drainage for both the wetting LIS and partially wetting LIS is shown in Fig.~\ref{fig:perc_lubricant_drainage}. After two hours, the wetting LIS (dark red markers) is completely drained ($\%LD=100$). Within the first 25 minutes, approximately 90\% of the lubricant volume is lost due to shear-driven drainage. This is followed by slower drainage of a thin film at the walls of the groove. For all partially wetting lubricants in longitudinal grooves without transversal barriers (indicated by the red, dark orange, and orange markers), finite retention is observed ($\%LD<100$). As anticipated by the presented theory,  lubricant depletion increases with the shear stress. There is a decay within the first 10 minutes caused by the disruption of the liquid-liquid interface induced by turbulent fluctuations. This allows water to penetrate and partially displace the lubricant in the grooves, resulting in the loss of lubricant droplets entrained in the bulk. The remaining lubricant forms elongated droplets that remain essentially stationary. Partially wetting lubricants in cavities (depicted by the green symbols) exhibit a considerably slower rate of drainage. The accompanying supplementary movies (Movie S1-Movie S7) demonstrate a very slow movement of partially wetting configurations and a slow change in fluorescence intensity. Both these effects contribute to $\%LD$ not fully saturating. 

\section{Analytical description of flow field}\label{app:analytics}

The Stokes equations are solved analytically to describe the fluid motion in the lubricant. 
We consider a completely oil-filled groove subjected to a uniform (constant) fluid-fluid interface shear stress $\tau_s$. The wall-normal coordinate $y = 0$ is at the bottom, while the spanwise coordinate $z = 0$ is at the centerline of the groove.

\subsection{Problem definition}
Following  Wexler et al. \cite{Wexler2015RobustWettability}, it is assumed that the flow is unidirectional, with the streamwise velocity $u = u(y,z)$ satisfying
\begin{equation}
    \mu_l \nabla^2 u = \frac{\text{d} p}{\text{d} x},
    \label{eq:momentumEquation}
\end{equation}
where $\text{d} p/\text{d} x$ is a constant pressure gradient inside the groove. The boundary conditions are
\begin{subequations}
    \begin{align}
        u &= 0  \quad &\text{for} \quad z &= \pm w/2 &\text{ (side walls)}, \\
        u &= 0  \quad &\text{for} \quad y &= 0 &\text{ (bottom wall)}, \\
        \mu_l\frac{\partial u}{\partial y} &= \tau_s \quad &\text{for} \quad y &= k &\text{ (top boundary)}.
    \end{align}
\end{subequations}
Since Eq.~\ref{eq:momentumEquation} is linear, we can consider $u$ as a superposition of two solutions $u_s = u_s(y,z)$ and $u_p = u_p(y,z)$, driven by the imposed shear stress and the pressure gradient, respectively. They satisfy
\begin{equation}
    \mu_l \nabla^2 u_s = 0,
    \label{eq:momentumEquationS}
\end{equation}
\begin{subequations}
    \begin{align}
        u_s &= 0  \quad &\text{for} \quad z &= \pm w/2 &\text{ (side walls)}, \label{eq:BCSideS}\\
        u_s &= 0  \quad &\text{for} \quad y &= 0 &\text{ (bottom wall)}, \label{eq:BCBottomS} \\
        \mu_l\frac{\partial u_s}{\partial y} &= \tau_s \quad &\text{for} \quad y &= k &\text{ (top boundary)}, \label{eq:BCTopS}
    \end{align}
\end{subequations}
and
\begin{equation}
    \mu_l \nabla^2 u_p = \frac{\text{d} p}{\text{d} x},
    \label{eq:momentumEquationP}
\end{equation}
\begin{subequations}
    \begin{align}
        u_p &= 0  \quad &\text{for} \quad z &= \pm w/2 &\text{ (side walls)}, \label{eq:BCSideP}\\
        u_p &= 0  \quad &\text{for} \quad y &= 0 &\text{ (bottom wall)}, \label{eq:BCBottomP}\\
        \mu_l\frac{\partial u_p}{\partial y} &= 0 \quad &\text{for} \quad y &= k &\text{ (top boundary)}. \label{eq:BCTopP}
    \end{align}
\end{subequations}

\subsection{Solution by separation of variables}
The flow field can be found by separation of variables and eigenfunction expansions \citep{shah78, Wexler2015RobustWettability}.

\subsubsection{Shear-driven flow}
Starting with $u_s$, we assume
\begin{equation}
    u_s(y,z) = Y_s(y)Z_s(z).
\end{equation}
Eq.~\ref{eq:momentumEquationS} implies
\begin{equation}
    Y_s''Z_s + Y_sZ_s'' = 0 \implies \frac{Y_s''}{Y_s} = -\frac{Z_s''}{Z_s} = \lambda,
\end{equation}
where $\lambda$ is a constant. We consider $\lambda > 0$, for which the solution for $Z_s$ is
\begin{equation}
    Z_s = a\cos\left(\sqrt{\lambda} z\right) + b\sin\left(\sqrt{\lambda} z\right),
\end{equation}
for coefficients $a$ and $b$. We are only interested in solutions symmetric with respect to $z = 0$, giving $b = 0$. The boundary conditions \eqref{eq:BCSideS} equal $Z_s(w/2) = Z_s(-w/2) = 0$, giving $\sqrt{\lambda} = \pi (2n+1)/w$, where $n = 0,1,2,\dots$~. The solution for $Y_s$ can be written
\begin{equation}
    Y_s = c\cosh\left(\frac{\pi \left (2n + 1 \right)}{w} y\right) + d\sinh\left(\frac{\pi (2n + 1)}{w} y\right),
    \label{eq:usGeneralYSolution}
\end{equation}
for coefficients $c$ and $d$. The boundary condition \eqref{eq:BCBottomS} implies $Y_s(0) = 0$, giving $c = 0$. An expression for $u_s$ is therefore
\begin{equation}
      u_s = \sum_{n = 0}^\infty d_n\sinh\left(\frac{\pi (2n + 1)}{w} y\right) \cos\left(\frac{\pi (2n + 1)}{w}z\right),
      \label{eq:usGeneralSolution}
\end{equation}
for coefficients $d_n$.

The inhomogeneous boundary condition \eqref{eq:BCTopS} must also be satisfied. First, we take the derivative of Eq.~\ref{eq:usGeneralSolution}, to obtain,
\begin{equation}
    \mu_l\frac{\partial u_s}{\partial y} = \mu_l\sum_{n = 0}^\infty d_n\frac{\pi (2n + 1)}{w}\cosh\left(\frac{\pi (2n + 1)}{w} y\right) \cos\left(\frac{\pi (2n + 1)}{w}z\right).
    \label{eq:usStress}
\end{equation}
Then, we expand the constant function over the interval $(-w/2, w/2)$ in a Fourier series as (see e.g.~\citep{rade99}) 
\begin{equation}
    \tau_s = \frac{4\tau_s}{\pi}\sum_{n = 1}^\infty \frac{\sin(n\pi/2)}{n}\cos\left(\frac{\pi n}{w}z\right) = \frac{4\tau_s}{\pi}\sum_{n = 0}^\infty \frac{(-1)^n}{2n+1}\cos\left(\frac{\pi(2n+1)}{w}z\right)
    \label{eq:shearStressExpansion}
\end{equation}
assuming that its wavelength is $2w$. Matching Eqs.~\ref{eq:usStress} and \ref{eq:shearStressExpansion}, we have
\begin{equation}
    d_n = \frac{4w\tau_s}{\mu_l\pi^2}\frac{(-1)^n}{(2n+1)^2}\left[\cosh\left(\frac{\pi (2n + 1)}{w} k\right)\right]^{-1},
\end{equation}
and thus,
\begin{equation}
      u_s = \frac{4w\tau_s}{\mu_l}\sum_{n = 0}^\infty \frac{(-1)^n}{\pi^2(2n+1)^2}\frac{\sinh\left(\frac{\pi (2n + 1)}{w} y\right)}{\cosh\left(\frac{\pi (2n + 1)}{w} k\right)} \cos\left(\frac{\pi (2n + 1)}{w}z\right).
      \label{eq:SI-us}
\end{equation}
To compute the flux, we integrate over the cross-section. We have that
\begin{equation}
    \int_{-w/2}^{w/2} \cos\left(\frac{\pi (2n + 1)}{w}z\right) \text{d} z = \frac{2w}{\pi (2n + 1)}\sin\left(\frac{\pi}{2} (2n + 1)\right) = \frac{2w}{\pi (2n + 1)}(-1)^n
    \label{eq:cosIntegral}
\end{equation}
and
\begin{equation}
    \int_{0}^{k} \sinh\left(\frac{\pi (2n + 1)}{w} y\right) \text{d} y = \frac{w}{\pi (2n + 1)}\left(\cosh\left(\frac{\pi (2n + 1)}{w} k\right) - 1 \right).
    \label{eq:sinhIntegral}
\end{equation}
Hence,
\begin{eqnarray}\nonumber
    q_{\tau} &=& \int_{-w/2}^{w/2}\int_{0}^{k}u_s \text{d} y\text{d} z  \\\nonumber
   &=& \frac{w^3\tau_s}{\mu_l}\sum_{n = 0}^\infty \frac{8}{(2n+1)^4\pi^4}\left(1 - \frac{1}{\cosh\left(\frac{\pi (2n + 1)}{w} k\right)}\right) \\\nonumber
    &=& \frac{w^3\tau_s}{\mu_l}\left(\frac{1}{12} - \sum_{n = 0}^\infty \frac{8}{(2n+1)^4\pi^4}\frac{1}{\cosh\left(\frac{\pi (2n + 1)}{w} k\right)}\right) \\ \nonumber
    &=& \frac{wk^2\tau_s}{\mu_l}\left(\frac{1}{12}\frac{w^2}{k^2} - \frac{w^2}{2k^2}\sum_{n = 0}^\infty \frac{1}{(n+1/2)^4\pi^4}\frac{1}{\cosh\left(\frac{\pi(n + 1/2)}{w} 2k\right)}\right) \\
    &=& \frac{wk^2\tau_s}{\mu_l}c_s.
    \label{eq:shearDrivenFlux}
\end{eqnarray}
In the above expression, we used that (see e.g.~\cite{rade99}) 
\begin{equation}
    \sum_{n = 0}^\infty \frac{1}{(2n+1)^4\pi^4} = \frac{\pi^4}{96}.
\end{equation}
We also introduced the geometrical resistance constant $c_s$ as a function of $w/k$. An alternative, but equivalent, expression for $c_s$ was given by \cite{Wexler2015RobustWettability},
\begin{equation}
    c_s = \frac{1}{2} - \frac{4k}{w}\sum_{n = 0}^\infty \frac{(-1)^n}{(n+1/2)^4\pi^4}\tanh\left(\frac{\pi(n + 1/2)}{2k} w\right).
\end{equation}

\begin{figure}
    \centering
    \includegraphics[width=0.8\linewidth]{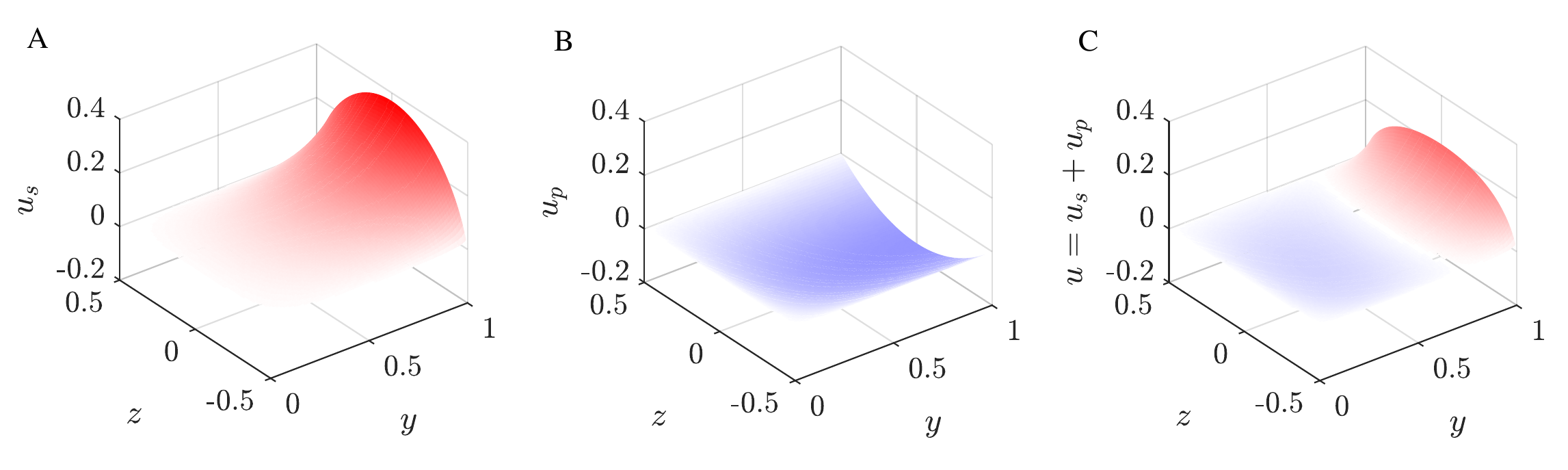}
    \caption{
   Analytical flow fields for groove width ($w/k=1$) shown for (A) shear-driven velocity ($u_s$), (B) pressure-gradient-driven velocity ($u_p$), and (C) total velocity ($u = u_s + u_p$). Coordinates are normalized by $k$. The solutions were obtained by evaluating equations \eqref{eq:SI-us} and \eqref{eq:SI-up}. The shear-driven solution ($u_s$) in (A) generates a positive flux and exhibits its highest magnitude at the groove center ($z=0$) and at the boundary where the shear stress is imposed ($y=1$). The pressure-gradient-driven solution ($u_p$) in (B) induces a negative flux and has its maximum magnitude at the groove center ($z=0, y=0.5k$). The net flow (C) has zero flux when integrated over the $y$ and $z$ coordinates. } 
    \label{fig:analyticalFlowFields}
\end{figure}

\subsubsection{Pressure-driven flow}
We now move on to finding $u_p$ for  the inhomogeneous system given Eq.~\ref{eq:momentumEquationP}. We assume a solution similar to Eq.~\ref{eq:usGeneralSolution},
\begin{equation}
    u_p = \sum_{n = 0}^\infty f_n(y) \cos\left(\frac{\pi (2n + 1)}{w}z\right),
\end{equation}
with functions $f_n(y)$. Similar as for $\tau_s$ in Eq.~\ref{eq:shearStressExpansion}, we can write 
\begin{equation}
    \frac{\text{d} p}{\text{d} x} = \frac{\text{d} p}{\text{d} x}\frac{4}{\pi}\sum_{n = 1}^\infty \frac{\sin(n\pi/2)}{n}\cos\left(\frac{\pi n}{w}z\right) = \frac{\text{d} p}{\text{d} x}\frac{4}{\pi}\sum_{n = 0}^\infty \frac{(-1)^n}{2n+1}\cos\left(\frac{\pi(2n+1)}{w}z\right).
    \label{pressureExpansion}
\end{equation}
Eq.~\ref{eq:momentumEquationP} then results in
\begin{equation}
\sum_{n = 0}^\infty \left(f_n''(y) - f_n(y)\frac{\pi^2 (2n + 1)^2}{w^2}\right) \cos\left(\frac{\pi (2n + 1)}{w}z\right) \\ = \frac{1}{\mu_l}\frac{\text{d} p}{\text{d} x}\frac{4}{\pi}\sum_{n = 0}^\infty \frac{(-1)^n}{2n+1}\cos\left(\frac{\pi(2n+1)}{w}z\right).
\end{equation}
The general solutions can be found by adding a particular solution and a homogeneous solution to 
fulfil the boundary conditions,
\begin{equation}
    u_p = u_p^\textit{inhom} + u_p^\textit{homog}.
\end{equation}
A particular solution is simply given by
\begin{equation}
    f_n = -\frac{w^2}{\mu_l}\frac{\text{d} p}{\text{d} x}\frac{4(-1)^n}{(2n+1)^3\pi^3} \\ \implies u_p^\textit{inhom}= -\sum_{n = 0}^\infty \frac{w^2}{\mu_l}\frac{\text{d} p}{\text{d} x}\frac{4(-1)^n}{(2n+1)^3\pi^3} \cos\left(\frac{\pi (2n + 1)}{w}z\right).
\end{equation}
The general solution to the homogeneous equation is equivalent to Eq.~\ref{eq:usGeneralSolution}, so that
\begin{eqnarray}\nonumber
    u_p &=& u_p^\textit{inhom} + u_p^\textit{homog} \\
    &=& \sum_{n = 0}^\infty\left(c_n\cosh\left(\frac{\pi (2n + 1)}{w} y\right)\right. + d_n\sinh\left(\frac{\pi (2n + 1)}{w} y\right) - \left.\frac{w^2}{\mu_l}\frac{\text{d} p}{\text{d} x}\frac{4(-1)^n}{(2n+1)^3\pi^3} \right)\cos\left(\frac{\pi (2n + 1)}{w}z\right), 
    \label{eq:SI-up}
\end{eqnarray}
for coefficients $c_n$ and $d_n$. The boundary conditions on the side walls are fulfilled (eq.~\ref{eq:BCSideP}). The bottom wall no-slip condition \eqref{eq:BCBottomP} and the top wall no-shear condition \eqref{eq:BCTopP} imply, \begin{equation}
    c_n = \frac{w^2}{\mu_l}\frac{\text{d} p}{\text{d} x}\frac{4(-1)^n}{(2n+1)^3\pi^3} \quad \text{ and } \quad d_n = -c_n\tanh\left(\frac{\pi (2n + 1)}{w} k\right),
\end{equation}
respectively. The complete solution is
\begin{eqnarray}\nonumber
    u_p &=& 
    -\frac{w^2}{\mu_l}\frac{\text{d} p}{\text{d} x}\sum_{n = 0}^\infty\frac{4(-1)^n}{(2n+1)^3\pi^3}\left(1 - \cosh\left(\frac{\pi (2n + 1)}{w} y\right)\right.\\
    &&+ \left.\tanh\left(\frac{\pi (2n + 1)}{w} k\right)\sinh\left(\frac{\pi (2n + 1)}{w} y\right)\right)\cos\left(\frac{\pi (2n + 1)}{w}z\right).
\end{eqnarray}
To evaluate the corresponding flux, we use that the integral
\begin{equation}
    \int_{0}^{k} \cosh\left(\frac{\pi (2n + 1)}{w} y\right) \text{d} y = \frac{w}{\pi (2n + 1)}\sinh\left(\frac{\pi (2n + 1)}{w} k\right).
    \label{eq:coshIntegral}
\end{equation}
Eqs.~\ref{eq:cosIntegral}, \ref{eq:sinhIntegral}, and \ref{eq:coshIntegral} give the flux
\begin{eqnarray}\nonumber
    q_p = \int_{-w/2}^{w/2}\int_{0}^{k}u_p \text{d} y\text{d} z &=&
    -\frac{w^3}{\mu_l}\frac{\text{d} p}{\text{d} x}\sum_{n = 0}^\infty\frac{8}{(2n+1)^4\pi^4}\left(h - \frac{w}{\pi (2n + 1)}\sinh\left(\frac{\pi (2n + 1)}{w} k\right)\right. \\\nonumber
    &&+ \left.\tanh\left(\frac{\pi (2n + 1)}{w} k\right)\frac{w}{\pi (2n + 1)}\left(\cosh\left(\frac{\pi (2n + 1)}{w} k\right) - 1 \right)\right) \\\nonumber
    &=& -\frac{w^3k}{\mu_l}\frac{\text{d} p}{\text{d} x}\left(\frac{1}{12} - \frac{w}{k}\sum_{n = 0}^\infty\frac{8}{(2n+1)^5\pi^5}\tanh\left(\frac{\pi (2n + 1)}{w} k\right)\right) \\\nonumber
    &=& -\frac{wk^3}{\mu_l}\frac{\text{d} p}{\text{d} x}\left(\frac{1}{12}\frac{w^2}{k^2} - \frac{w^3}{4k^3}\sum_{n = 0}^\infty\frac{1}{(n+1/2)^5\pi^5}\tanh\left(\frac{\pi (n + 1/2)}{w} 2k\right)\right) \\
    &=& -c_p\frac{wk^3}{\mu_l}\frac{\text{d} p}{\text{d} x},
    \label{eq:pressureDrivenFlux}
\end{eqnarray}
where $c_p$ is a geometrical resistance constant. An alternative expression, but equivalent, for $c_p$ was given by Wexler et al. \cite{Wexler2015RobustWettability},
\begin{equation}
    c_p = \frac{1}{3} - \frac{4k}{w}\sum_{n = 0}^\infty \frac{1}{(n+1/2)^5\pi^5}\tanh\left(\frac{\pi(n + 1/2)}{2k} w\right).
\end{equation}

\subsection{Required pressure drop}
For there to be no drainage of groove liquid, the fluxes need to balance,
\begin{equation}
    q_{\tau} + q_p = 0.
\end{equation}
The expression for the fluxes, \eqref{eq:shearDrivenFlux} and \eqref{eq:pressureDrivenFlux}, result in
\begin{equation}
    \frac{\text{d} p}{\text{d} x} = \frac{\tau_s}{k}\frac{c_s}{c_p}.
    \label{eq:balancingPressure}
\end{equation}
If the pressure gradient is less than this value, there is drainage of liquid. An example of flow fields with balancing shear stress and pressure gradient are shown in Fig.~\ref{fig:analyticalFlowFields}. 

\subsection{Droplet hysteresis}
The adhesion force is defined as
\begin{equation}
    F_{cl} = \oint_\textit{cl} \gamma \left (\mathbf{n}_{cl}\cdot \mathbf{e}_x \right )  dl.
\end{equation}
Here, $\mathbf{n}_{cl}$ is a unit vector that is tangent to the liquid-lubricant interface and normal to the contact line. We may express this vector in terms of normals of the solid surface and interface.
The interfacial tension force per unit length at the contact line projected onto the solid is $\gamma\cos \theta$. The angle $\theta$ is the contact angle which also is the angle between the interface normal at the contact line and the solid normal ($\mathbf{n}_i$ and $\mathbf{n}$, respectively). The projected force is in the direction of the projected interface normal $\mathbf{n}_\textit{cl,proj}$ (where $\mathbf{n}_\textit{cl}$ points out from the droplet). The net contribution from the surface tension force acting on the droplet at the contact line in the streamwise direction is, therefore
\begin{equation}
    F_{\textit{cl}} = \oint_\textit{contact line} \gamma\cos \theta\mathbf{n}_\textit{cl,proj}\cdot \mathbf{e}_x \text{d} s,
\end{equation}
where $\mathbf{e}_x$ is the streamwise unit vector. We assume that $\theta = \theta_\textit{rec}$ on the upstream side of the droplet and $\theta = \theta_\textit{adv}$ on the downstream, which are the simplest possible assumptions. Downstream and upstream contact lines at the bottom of the groove are perfectly perpendicular to the streamwise flow ($\mathbf{n}_\textit{cl,proj} = \pm \mathbf{e}_x)$ give a contribution
\begin{equation}
    -\gamma w(\cos \theta_\textit{rec} - \cos \theta_\textit{adv}),
\end{equation}
which is used here. We assume that the contact line on the walls goes all the way from the bottom to the top of the grooves ($y = 0$ to $y = k$). On the side walls, $\mathbf{n}_\textit{cl,proj}\cdot \mathbf{e}_x$ = $\pm\hat{\mathbf{s}}\cdot \mathbf{e}_y$, where $\hat{\mathbf{s}}$ is the contact line tangent unit vector in the direction of integration. Looking at the wall on the dowstream side where the integral starts from the bottom of the groove and goes to the top ($\mathbf{n}_\textit{cl,proj}\cdot \mathbf{e}_x$ = $-\hat{\mathbf{s}}\cdot \mathbf{e}_y$),
\begin{eqnarray}\nonumber
    \int_\textit{side wall cl} \gamma\cos \theta_\textit{rec}\mathbf{n}_\textit{cl,proj}\cdot \mathbf{e}_x \text{d} s &=& -\gamma\cos \theta_\textit{rec}\int_\textit{side wall cl} \mathbf{e}_y\cdot\hat{\mathbf{s}} \text{d} s \\\nonumber
   & =& -\gamma\cos \theta_\textit{rec}\int_\textit{side wall cl} \left(\left(\frac{\partial}{\partial x},\frac{\partial}{\partial y} \right)y\right)  \cdot \text{d} \mathbf{s} \\
   &=& -\gamma\cos \theta_\textit{rec}[y]_0^k = -\gamma k\cos \theta_\textit{rec}
\end{eqnarray}
by the gradient theorem. The forces from the other side-wall parts of the contact line can be computed in analog. The total contribution is
\begin{equation}
    -2\gamma k(\cos \theta_\textit{rec} - \cos \theta_\textit{adv}).
\end{equation}
Hence,
\begin{equation}
    F_{\textit{cl},x} = -(w + 2k)\gamma(\cos \theta_\textit{rec} - \cos \theta_\textit{adv}).
\end{equation}
This net force acting in the negative streamwise direction is assumed to give rise to interface curvature and a corresponding pressure gradient $\text{d} p/\text{d} x$ over the droplet. The reference pressure $p_0$ of the external flow is assumed to be constant, and the pressure differences over the interface at the upstream and downstream part ($p_\textit{upstr} - p_0$ and $p_\textit{downstr} - p_0$, respectively) are given by the curvature. With $F_{\textit{cl},x}$ assumed to change the pressure equally over the projected area $wk$ in the $yz$-plane, the effective pressure gradient can be written
\begin{equation}
    \frac{\text{d} p}{\text{d} x} = \frac{p_\textit{upstr} - p_0 - (p_\textit{downstr} - p_0)}{L} = \frac{p_\textit{upstr} -p_\textit{downstr}}{L} = -\frac{F_{\textit{cl},x}}{Lwk},
\end{equation}
for a droplet length $L$. Together with the flux balance \eqref{eq:balancingPressure}, the maximum stationary-state droplet length becomes
\begin{equation}
    L_\infty = \frac{\gamma}{\tau_s w}(\cos \theta_\textit{rec} - \cos \theta_\textit{adv})(w + 2k)\frac{c_p}{c_s}. 
    \label{eq:maximumRetentionLength}
\end{equation}

\section{Numerical model}\label{app_numerics}
\begin{figure}[tbhp]
\centering
\includegraphics[width=0.4\linewidth]{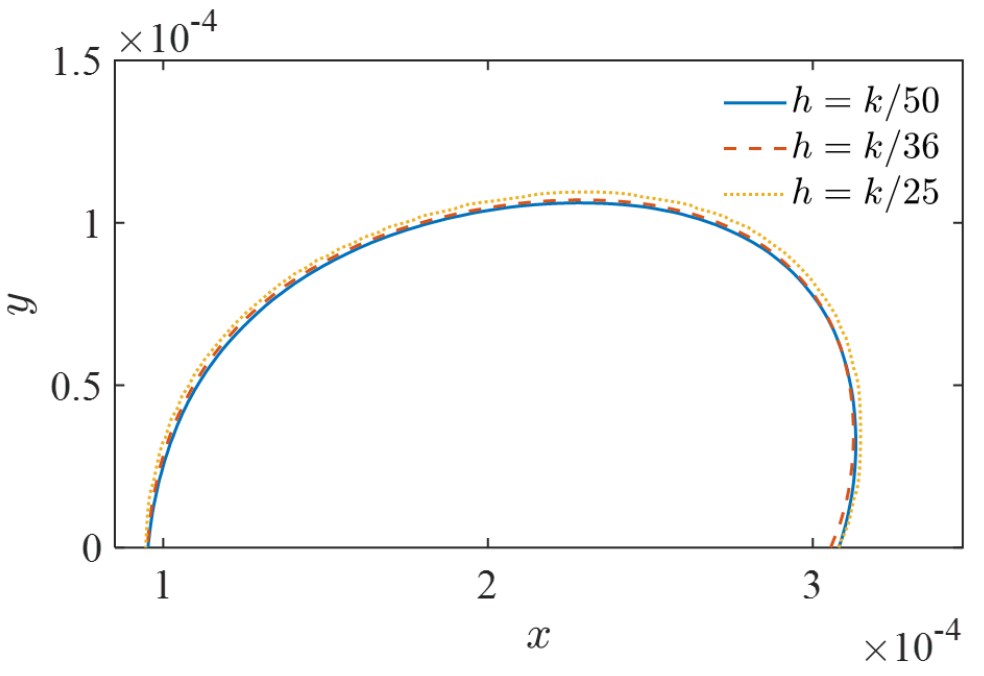}
\caption{Comparison of the equilibrium interface profiles at the centerline of the groove for three different grid resolutions. Streamwise $(x)$ and wall normal ($y$) coordinates are in meters. Here, $h$ denotes the grid-cell size and $k$ is groove height. We observe that the curvature of lubricant-liquid interface of the stationary droplet exhibits very small changes with respect to the resolution.
The parameters  $\theta_{\mathrm{adv}}=120^{\circ}$, $\theta_{\mathrm{rec}}=80^{\circ}$, $\tau=\SI{111}{\pascal}$, and $L=\SI{220}{\mu\meter}$ were kept constant, as well as computational parameters including the Courant–Friedrichs–Lewy condition, $\text{CFL}=0.5$. 
We also calculated the relative error $err = {|l-\hat{l}|}/{\hat{l}}$, with respect to $h=k/50$, 
where $l$ is the interfacial length for $h/25$ or $h/36$, and $\hat{l}$ is is the interfacial length for $h/50$. 
The resulting $err$ values for $h/25$ and $h/36$ are $2.6\%$ and $0.70\%$, respectively, suggesting that the grid size $h/36$ is sufficiently fine to produce accurate results.
}
\label{fig:grid_test}
\end{figure}

We consider a single lubricant droplet exposed to a fully developed shear flow. The flow domain has the size $\left(L_x, \ L_y, \ L_z\right)=\left(20k, \ w, \ 2k\right)$, where $x$, $y$, $z$ represent the streamwise, wall-normal and spanwise directions, respectively. We consider a half pitch in the spanwise direction -- i.e., from the centreline of a groove to the centerline of a crest -- and impose symmetry boundary conditions on both sides. Along the streamwise direction, a periodic boundary condition is imposed. At the top side, a moving wall boundary with a constant streamwise velocity $U_t$ is set. The streamwise-aligned groove has a rectangular cross-section, and its size ratio is $k/w=9/7$. 
For computational feasibility, the droplet length is limited to $15k$, which is shorter than the lengths observed in experiments. Therefore, we increase the shear stress ($\tau_s \in \left[33, \ 56, \ 111 \right] $ Pa) or decrease the hysteresis ($\Delta \theta  \in \left[10^{\circ}, \ 20^{\circ}, \ 30^{\circ}, \ 40^{\circ}, \ 50^{\circ}, \ 60^{\circ} \right] $)
to find the boundary between the regimes where droplets are stationary and moving. The numerical scheme in OpenFOAM employs a geometric VOF-based method for interface capturing \cite{gamet2020validation, scheufler2019accurate}. At solid walls, contact angle hysteresis is implemented by using the Robin boundary condition (see \cite{linder2015numerical} for details). Finally, the flow domain is meshed with a uniform cubic cell, and the chosen cell size $k/36$ has been confirmed to produce convergent results through a sensitivity analysis (Fig.~\ref{fig:grid_test}). The lubricant droplet reaches a steady state, i.e., stably moving or retaining, within  $40k/U_t$, and the final state of the droplet is recorded at $45k/U_t$.

\bibliography{bibliography}

\end{document}


\maketitle

\SItext

\section{List of movies}

The following list describes the movies recorded during the experiments.








\paragraph{Movie S1:} Wetting LIS in turbulent flow at $\tau_{max}=\SI{5.8}{\pascal}$
\paragraph{Movie S2:} Partially-wetting LIS in turbulent flow at $\tau_{max}=\SI{5.8}{\pascal}$
\paragraph{Movie S3:} Partially-wetting LIS in turbulent flow at $\tau_{max}=\SI{10.4}{\pascal}$
\paragraph{Movie S4:} Partially-wetting LIS in turbulent flow at $\tau_{max}=\SI{14.0}{\pascal}$
\paragraph{Movie S5:} Partially-wetting-cavity LIS in turbulent flow at $\tau_{max}=\SI{5.8}{\pascal}$
\paragraph{Movie S6:} Partially-wetting-cavity LIS in turbulent flow at $\tau_{max}=\SI{10.4}{\pascal}$
\paragraph{Movie S7:} Partially-wetting-cavity LIS in turbulent flow at $\tau_{max}=\SI{14.0}{\pascal}$

\bibliography{SI/SI_references}